\documentclass[11pt,twoside]{article}
\usepackage{fullpage,url} \def\Ungor{\"Ung\"or}  
\usepackage{graphicx}
\usepackage{multirow}
\usepackage{rotating}

\usepackage{amssymb,amsfonts,amsmath,amsthm}
\usepackage{latexsym}
\usepackage{epsfig}
\usepackage{psfig}

\def\sect#1{Section~\ref{sec:#1}}
\def\figr#1{Figure~\ref{fig:#1}}

\newtheorem{lemma}{Lemma}
\newtheorem{theorem}[lemma]{Theorem}
\newtheorem*{corollary}{Corollary}
\newtheorem{problem}{Problem}
\theoremstyle{definition}
\newtheorem*{definition}{Definition}

\newcommand{\comment}[1]{}

\begin{document}


\title{Tiling space and slabs with acute tetrahedra}

\author{David Eppstein$^1$
        \and
        John M. Sullivan$^2$
        \and
        Alper \Ungor$^3$}

\date{\footnotesize 
$^1$ Dept.~of Information and Computer Science, 
     Univ.~of California at Irvine, eppstein@ics.uci.edu\\
\vspace{.8ex} 
$^2$ Dept.~of Mathematics, Univ.~of Illinois at
Urbana-Champaign, jms@math.uiuc.edu\\
$^3$ Dept.~of Computer Science, Duke Univ., ungor@cs.duke.edu}

\maketitle

\begin{abstract}
We show it is possible to tile three-dimensional space using
  only tetrahedra with acute dihedral angles.
We present several constructions to achieve this, including one
in which all dihedral angles are less than $77.08^\circ$,
and another which tiles a slab in space.
\end{abstract}

\setlength{\baselineskip}{1.19\baselineskip}

\section{Problem definition}

Triangulations of two and three-dimensional domains 
  find numerious applications in scientific computing, computer
  graphics, solid modeling and medical imaging.  
Most of these applications impose a quality constraint on the elements
  of the triangulation. 
Among the most popular quality criteria for elements~\cite{BernE92}
  are the aspect ratio (circumradius over inradius), the minimum
  dihedral angle, and the radius-edge ratio (circumradius over shortest edge).
However, many other quality criteria have been considered, including
  maximum dihedral angle.
Bern {\em et al.}\ for instance, studied nonobtuse
  triangulations~\cite{BernCER95,BernEG94},
  where domains are meshed with simplices having no obtuse angles.
In this paper, we consider a slightly stronger quality constraint: 
  all the dihedral angles in the mesh are forced to be acute
  (strictly less than $90^\circ$).
Although acuteness seems only slightly stronger than nonobtuseness,
  this problem turns out to be considerably harder than 
  the nonobtuse triangulation problem, as we observe below in 
  \sect{background}.

\begin{definition}
An angle is {\em acute} if it is strictly less than
  a right angle ($\tfrac\pi2=90^\circ$).
A simplex is {\em acute} if all its (interior)
  dihedral angles are acute. 
A triangulation is {\em acute} if all of its simplices are acute.
\end{definition}

\begin{problem}
Given a domain $\Omega$, compute an acute triangulation of $\Omega$.
\label{pro:acute_meshing}
\end{problem}

There has been extensive work on the two-dimensional version of this problem,
  for the special cases where the domain $\Omega$ 
  is a triangle, square, quadrilateral, or a finite point set
  \cite{BernEG94,CassidyL80,Gardner60,Lindgren64,Maehara00,Manheimer60}. 
We review those results in \sect{background}.
In three-dimensional space, however, almost nothing
  has been known about acute triangulations before now.
To the best of our knowledge, even the following relaxed form of the problem, 
  where the the input domain is the entire space,
  had not been addressed in the literature.

\begin{problem}
Is it possible to tile three-dimensional Euclidean space using acute tetrahedra?
\label{pro:tiling}
\end{problem}

We present an affirmative answer to this question,
  by several different constructions.
The two-dimensional analog of this problem has a trivial positive answer:
  congruent copies of any single triangle will tile the plane. 
However, this idea does not extend to 3D, 
  as the regular tetrahedron (for instance) cannot tile space. 
All tetrahedra known to tile space have right
  angles, as further discussed in \sect{background}.

We started this research on acute triangulations 
  because of a method developed for space-time meshing which required
  an acute base mesh. 
This and our other motivations are discussed in \sect{motivation}. 
\sect{background} surveys previous research in acute triangulations.
\sect{acuteness} investigates what acuteness means for a 
  three-dimensional simplex and gives a comparison of acute and
  Delaunay triangulations.
Constructions tiling three-dimensional space, and hence solving
  Problem \ref{pro:tiling}, are given in \sect{constructions}.
The paper concludes in \sect{eval} with a quality assesment
  of these constructions. 

\section{Motivation} \label{sec:motivation}

We were originally motivated to study acute triangulations by the 
  space-time meshing algorithm of \Ungor\ and Sheffer \cite{UngorS02}.
This \emph{tent-pitcher} algorithm
  was designed to discretize space-time domains into
  meshes that obey a certain cone constraint, 
  which requires all faces in the mesh to have 
  smaller slopes than the cones that define the domain of influence 
  imposed by the numerical (engineering) problem.  
(For instance, we might require simply that all faces make at most
  a $45^\circ$ angle with the horizontal.)
Because there is then a well-defined direction of information flow
  across element boundaries, such meshes enable the use of very
  efficient element-by-element methods 
  (including space-time discontinuous Galerkin methods)
  to solve a wide variety of numerical problems, for instance in elastodynamics.
The tent-pitching algorithm starts with a space mesh of the two- or
  three-dimensional input domain and constructs the 
  space-time mesh using an advancing front approach. 
The algorithm is known to generate a valid space-time mesh if the 
  initial space mesh is an acute triangulation~\cite{UngorS02},
  but may fail if there is an obtuse angle or even a right angle.

Later, Erickson \emph{et al.}~\cite{EricksonGSU02} proposed an
  improved version of the tent-pitching algorithm.
By removing the acute angle requirement,
  the new space-time algorithm works over arbitrary spatial domains. 
However, there is a loss of efficiency (more elements are required)
  whenever there is a nonacute angle.

Thus the study of Problem \ref{pro:acute_meshing}
  is motivated by current space-time meshing algorithms.
But even a solution to Problem \ref{pro:tiling} is useful,
  since it leads to a better understanding
  of the acute triangulation problem for more general input domains,
  and it also finds some direct applications in mesh generation. 

Spacial tilings of high quality have been used for designing 
  meshing algorithms:
Fuchs~\cite{Fuchs98}, Field and Smith~\cite{Field86, FieldS91} 
  and Naylor~\cite{Naylor99} built meshes by overlaying standard tilings 
  onto the given polyhedral domain. 
They used tilings known at the time,
  such as Sommerville constructions (\figr{Sommerville}) 
  and subdivided cubes (\figr{cubes}), which we discuss in \sect{tilings}.
Their approach has three steps, illustrated in \figr{almost_regular}:
\begin{figure}[tb]
\begin{center} \begin{tabular}{c c c}
\psfig{figure=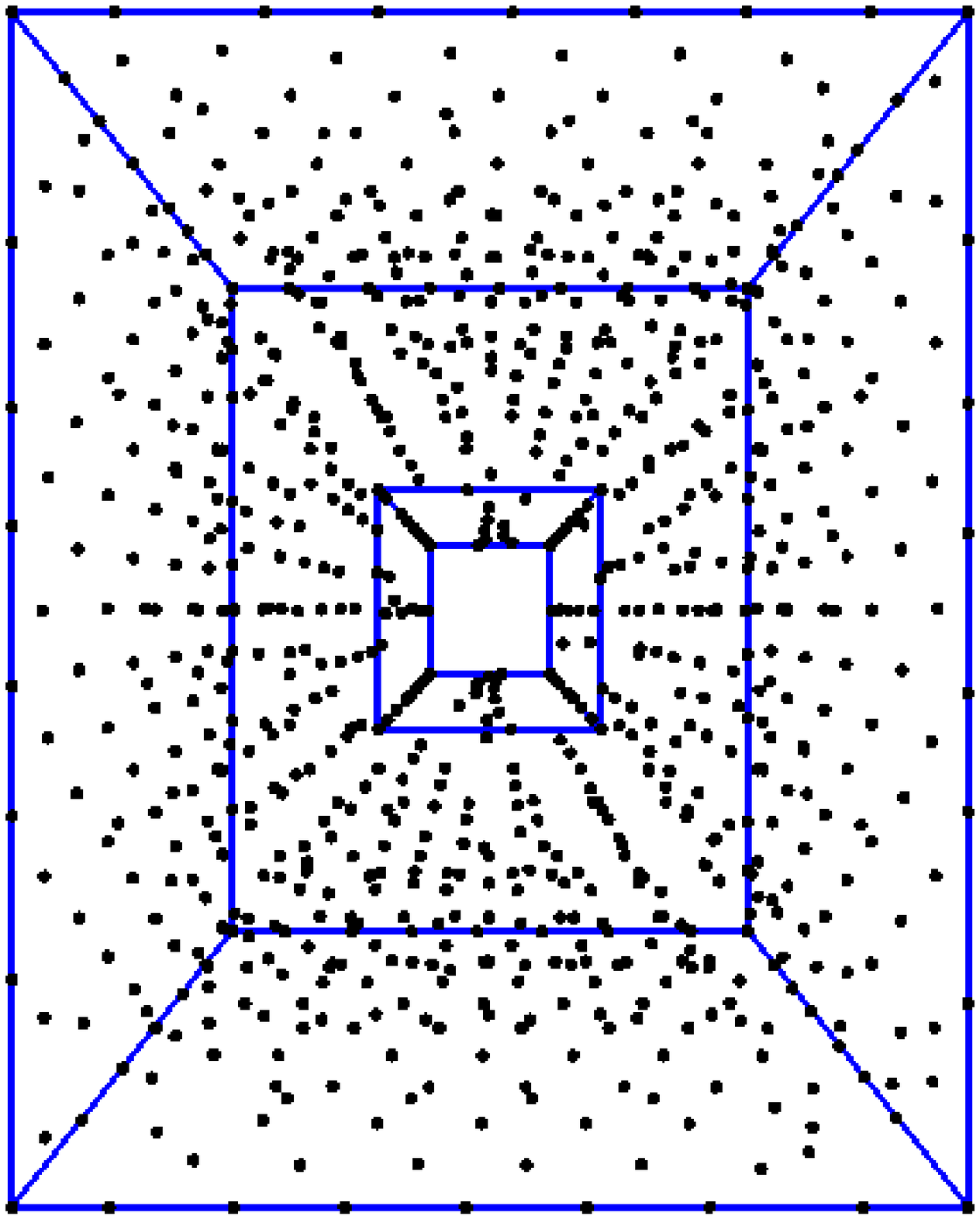,height=1.8in}&
\psfig{figure=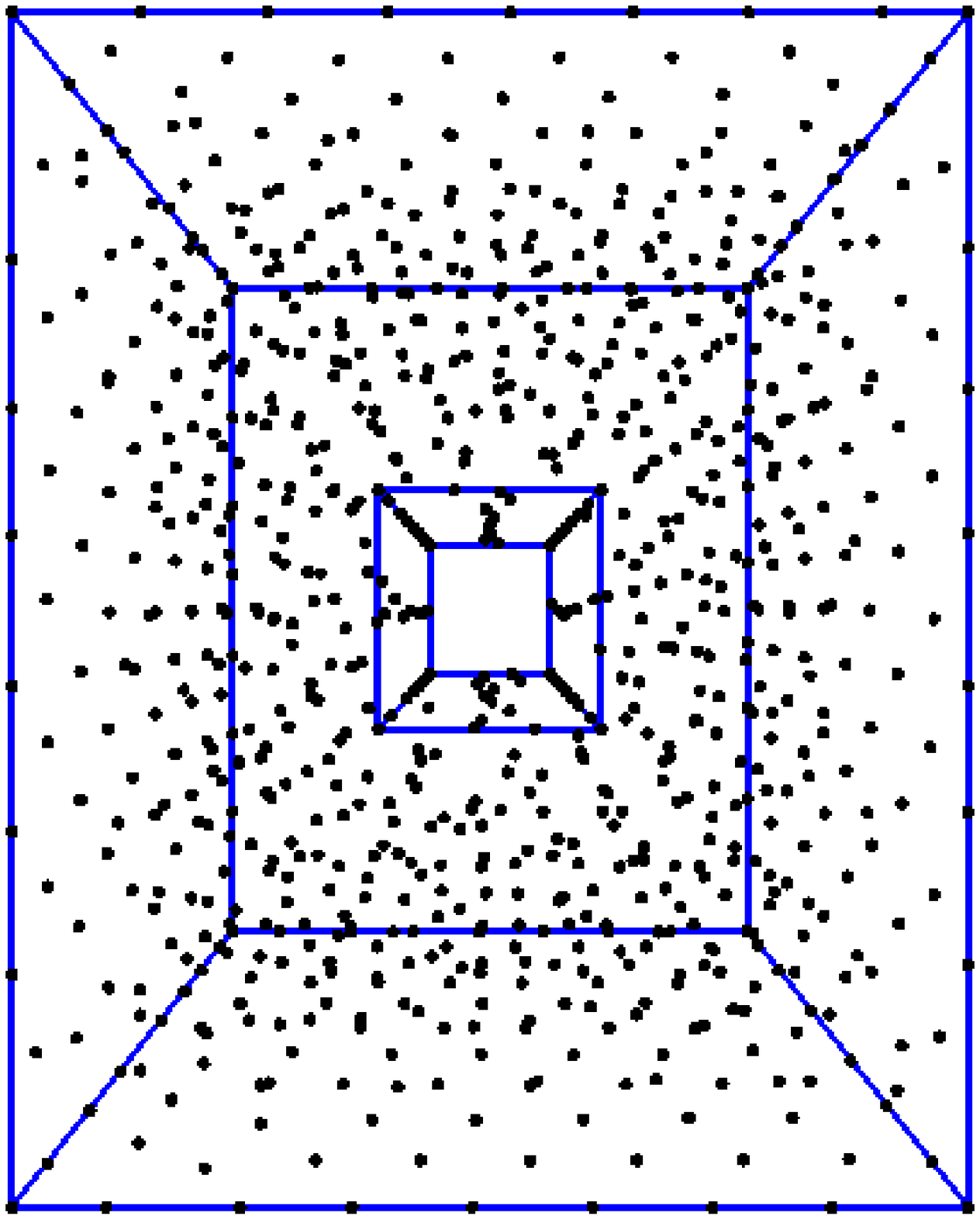,height=1.8in}&
\psfig{figure=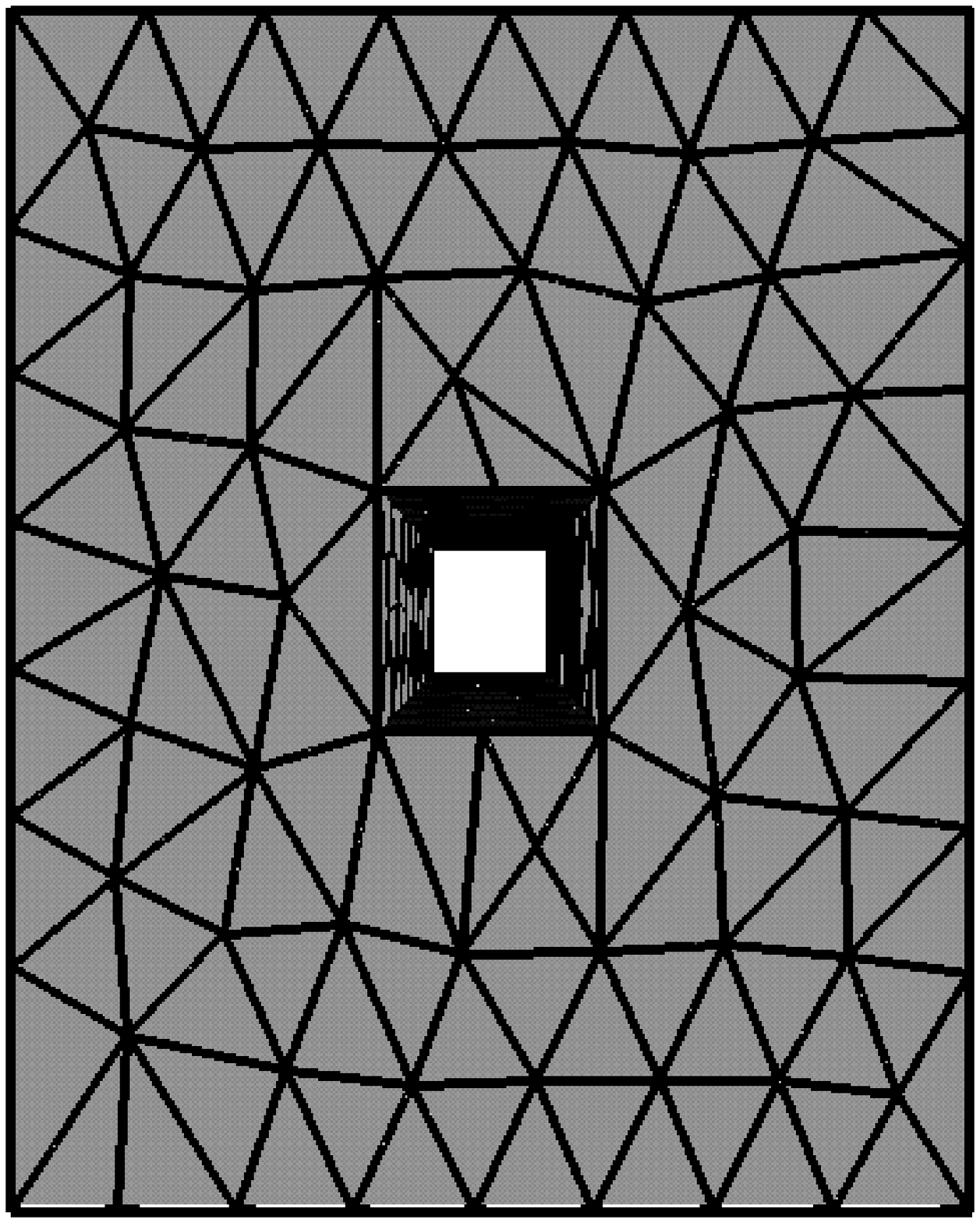,height=1.8in}\\
(a) & (b) & (c)\\
\end{tabular} \end{center}
\caption{An almost regular triangulation of a cube with a hole (A. Fuchs); 
(a) the point set of a body-centered cubic (BCC) lattice overlayed 
    with the domain; 
(b) the adjusted point set; (c) the conforming Delaunay triangulation.}
\label{fig:almost_regular}
\end{figure}

\begin{itemize}
\item[(a)] Overlay the chosen tiling with the given domain.
      The main challenge in this step is finding the right
      scaling, location and orientation for the tiling 
      so that it matches the domain boundary as closely as possible.
\item[(b)] Adjust the points to get a better fit.
      For this purpose, one of the standard smoothing techniques 
      \cite{CanannSB93,Field88,FreitagJP95} can be used.
      Alternatively, Fuchs \cite{Fuchs98} suggested minimizing a function
      which penalizes configurations that produce irregular vertices.
\item[(c)] Construct the mesh by computing the conforming Delaunay
      triangulation of the adjusted point set and the domain boundary.
\end{itemize}

Fuchs~\cite{Fuchs98} reports good performance of his experiments when
  he used the second Sommerville construction (\figr{Sommerville}(b))
  as the space tiling. 
(This tiling is the Delaunay triangulation of the body-centered cubic lattice.)
The dihedral angles of his mesh in \figr{almost_regular}(c)
  range between $7.6^\circ$ and $168.2^\circ$. 
However, most of the angles
  (here and also in his meshes of similar geometric domains)
  cluster around $60^\circ$ and $90^\circ$, which are exactly
  the dihedral angles of the BCC tetrahedron in the input tiling.
Some of the constructions we propose in \sect{constructions}
  are considerably better in terms of dihedral angles and also other quality
  measures. 
Our new constructions can find immediate use to improve the results
  of this previous research \cite{Fuchs98, Field86, FieldS91, Naylor99}
  on tiling-based meshing.

\comment{ 

}

Bossavit suggests~\cite{Bossavit01, Bossavit02}
  that acute triangulations may also be useful
  in computational electromagnetics.

\comment{
Kimmel and Sethian \cite{KimmelS98, Sethian99} also make use of acute
  triangulations in their work on computing geodesic paths on manifolds.
Their algorithm, called the fast marching method, computes geodesic 
  distances and shortest paths on triangulated domains. 
They provide a simpler version of the algorithm with a better accuracy 
  analysis when the underlying triangulation is acute.
}  

\section{Background} \label{sec:background}

\subsection{Acute and nonobtuse triangulations}

\begin{figure}[htb] \begin{center}
\begin{tabular}{c c c c}
\psfig{figure=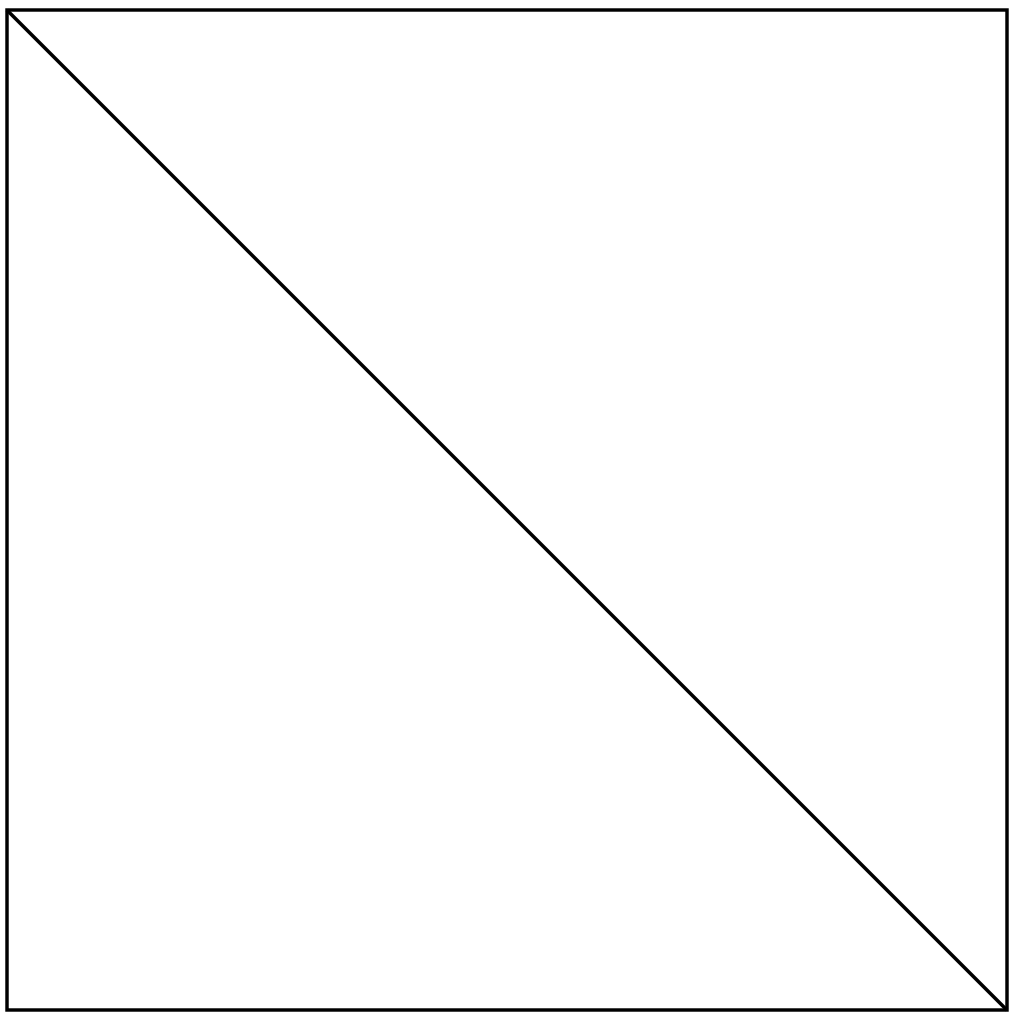,height=1.45in}&
\psfig{figure=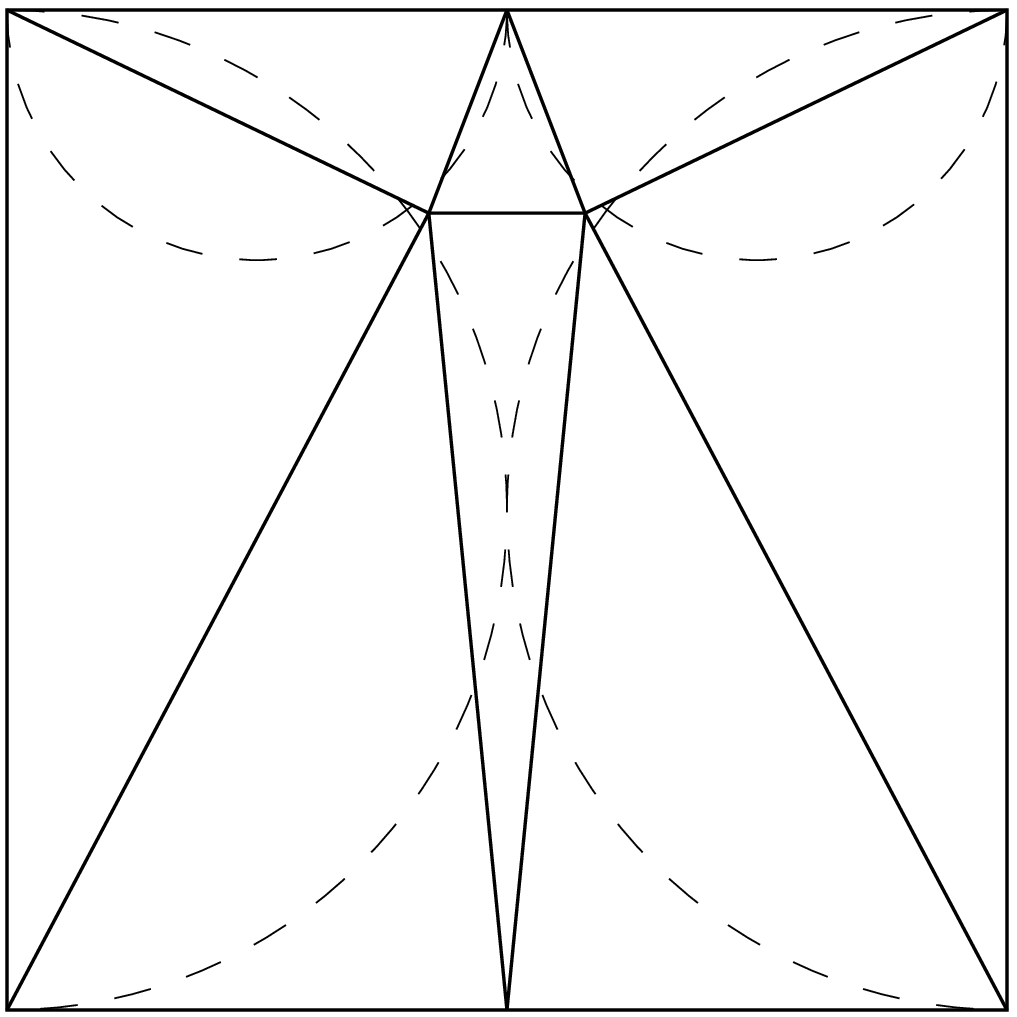,height=1.45in} &
\psfig{figure=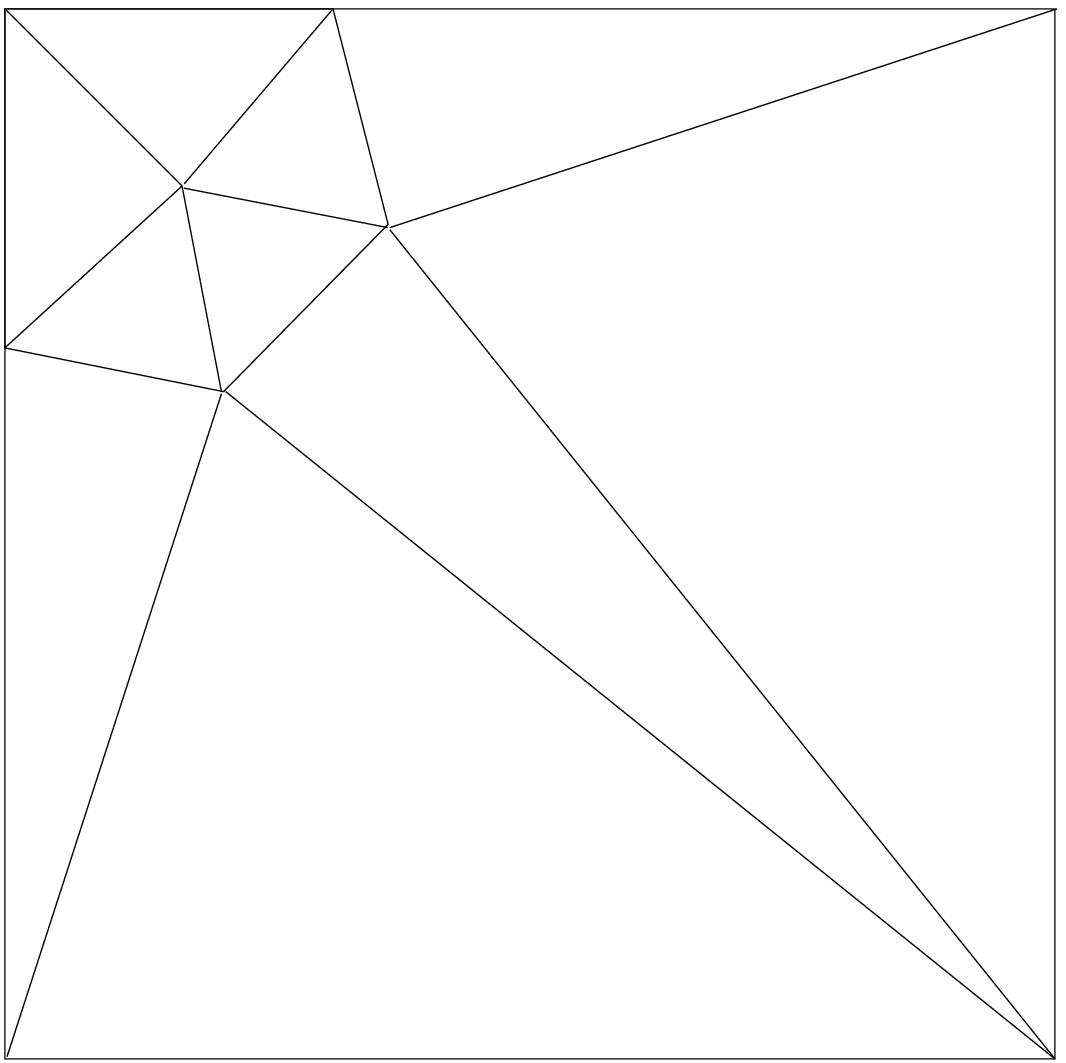,height=1.45in} &
\psfig{figure=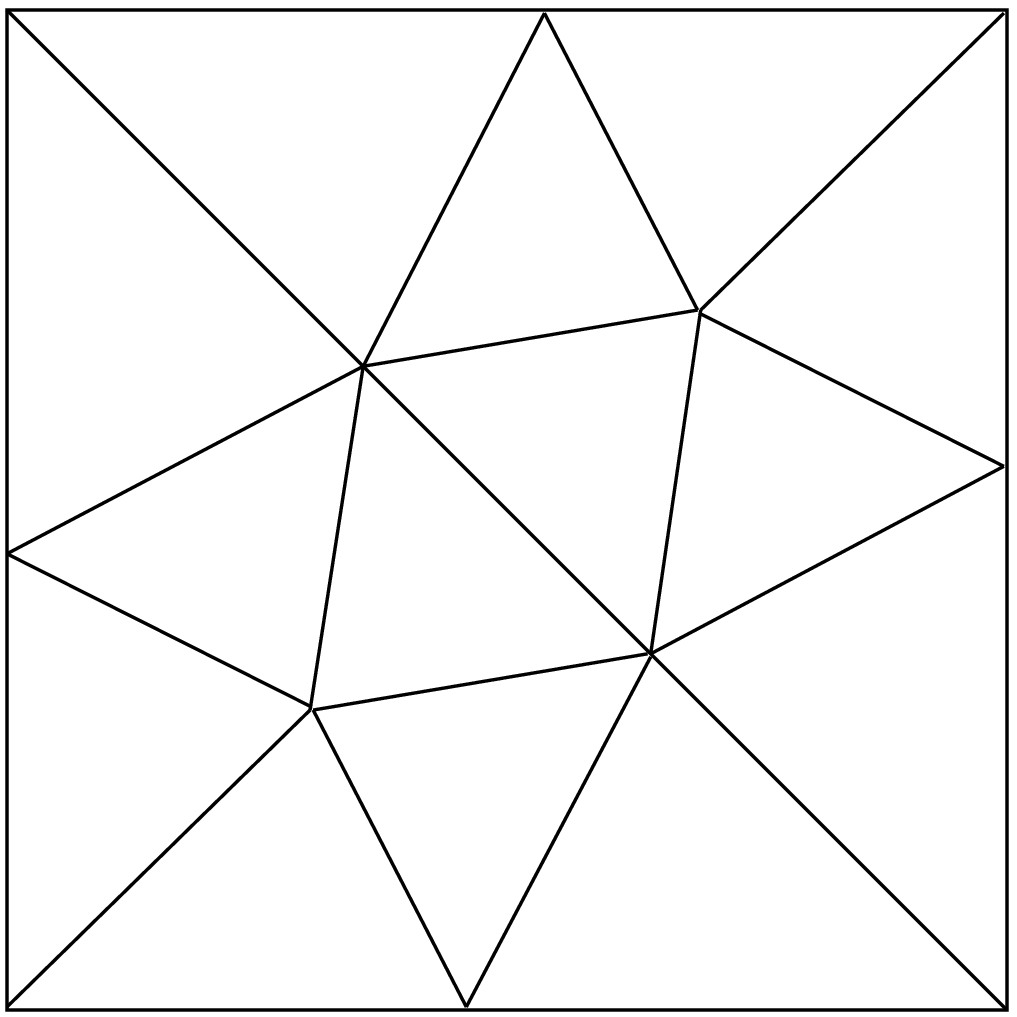,height=1.45in}\\
(a) & (b) & (c) & (d) \\
\end{tabular}
\caption{ (a) nonobtuse triangulation of a square 
(b) a square meshed with eight acute triangles 
(c) a square meshed with ten acute triangles 
(d) triangulation where maximum angle is $72^\circ$}
\label{fig:sq_acute}
\end{center} \end{figure}

There has been considerable research 
  \cite{BakerGR88, BernCER95, BernE92, BernMR94, Eppstein94}
  on the nonobtuse triangulation problem, which imposes a slightly
  weaker constraint than the acute triangulation problem.
Angles in a nonobtuse triangulation are less than or equal to $90^\circ$.
Bern {\em et al.}~\cite{BernCER95}
  showed that any $d$-dimensional point set can be 
  triangulated with $O(n^{\lceil d/2 \rceil})$ simplices, 
  none of which has any obtuse dihedral angles. 
However, they also proved that no similar result is possible
  if all angles are required to be at most $90^\circ - \epsilon$.
This indicates that the acute triangulation 
  problem is much more challenging than nonobtuse triangulation.
To appreciate this difference,
  consider the two problems for a square domain in two dimensions. 
A single diagonal cuts a square into two nonobtuse triangles,
  as in \figr{sq_acute}(a).
Finding an acute triangulation, however, can be a challenging
  recreational math problem. 

Lindgren~\cite{Lindgren64} showed that at least eight triangles,
  as in \figr{sq_acute}(b), are needed.
Later, Cassidy and Lord~\cite{CassidyL80} 
  showed that for any $n\ge10$ (but not for $n=9$)
  there is an acute triangulation with exactly $n$ triangles.
\figr{sq_acute}(c) shows the solution with ten triangles.
We can use the maximum angle in a triangulation as a quality measure.
The triangulations in \figr{sq_acute}(b,c) can be realized with
  maximum angles about $85^\circ$ and $80.3^\circ$, respectively.
Eppstein~\cite{Eppstein} improved this angle to  $72^\circ$
  using fourteen acute triangles, as shown in \figr{sq_acute}(d).
Using Euler's formula, Eppstein also showed that any
  acute triangulation of a square must have an interior vertex of
  valence five, implying that $72^\circ$ is the best possible. 
It is unknown whether there is a triangulation achieving this
  with fewer than fourteen triangles.

\begin{figure}[tb]
\begin{center}
\begin{tabular}{c c}
\psfig{figure=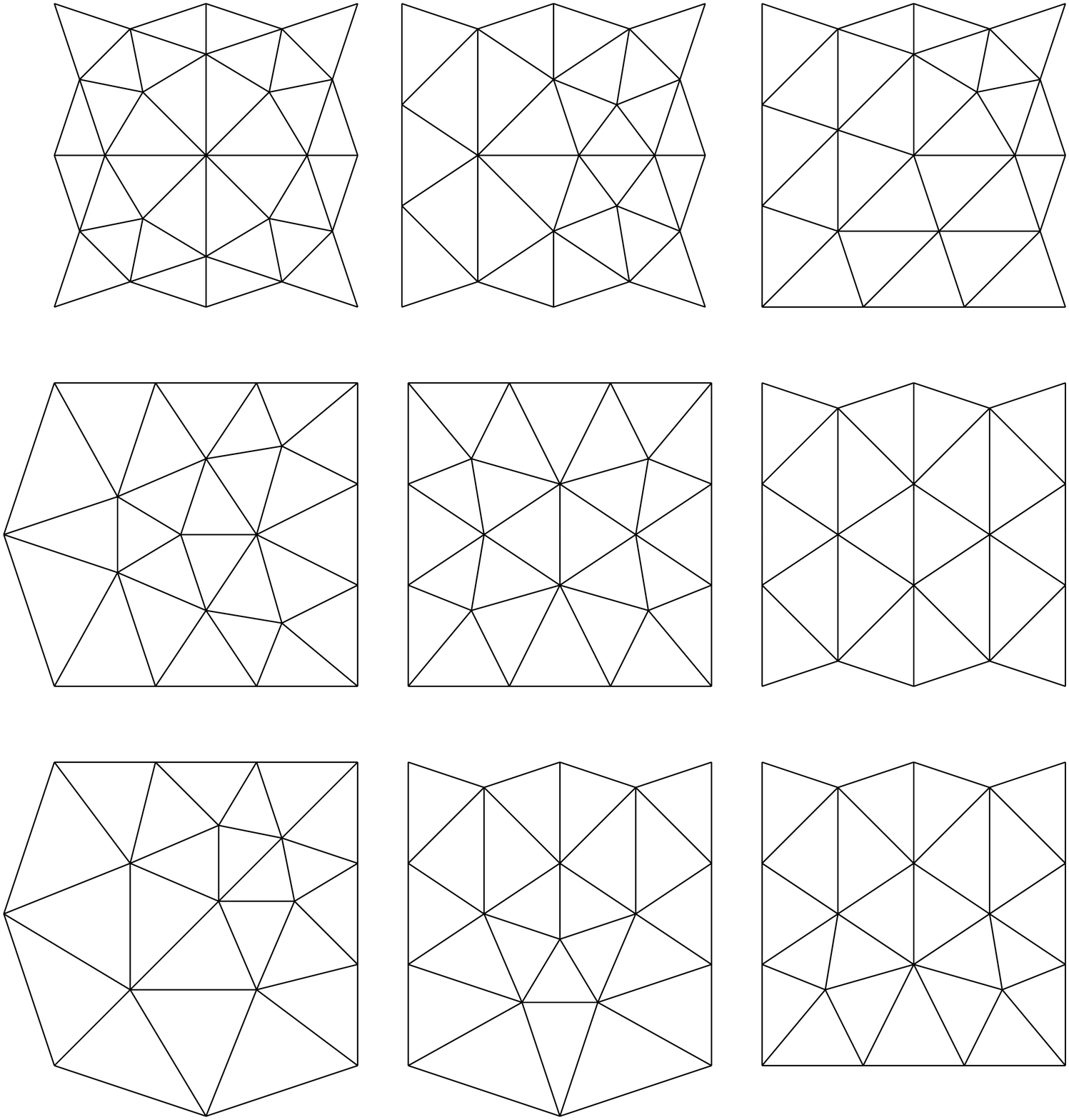,height=2.6in} &
\psfig{figure=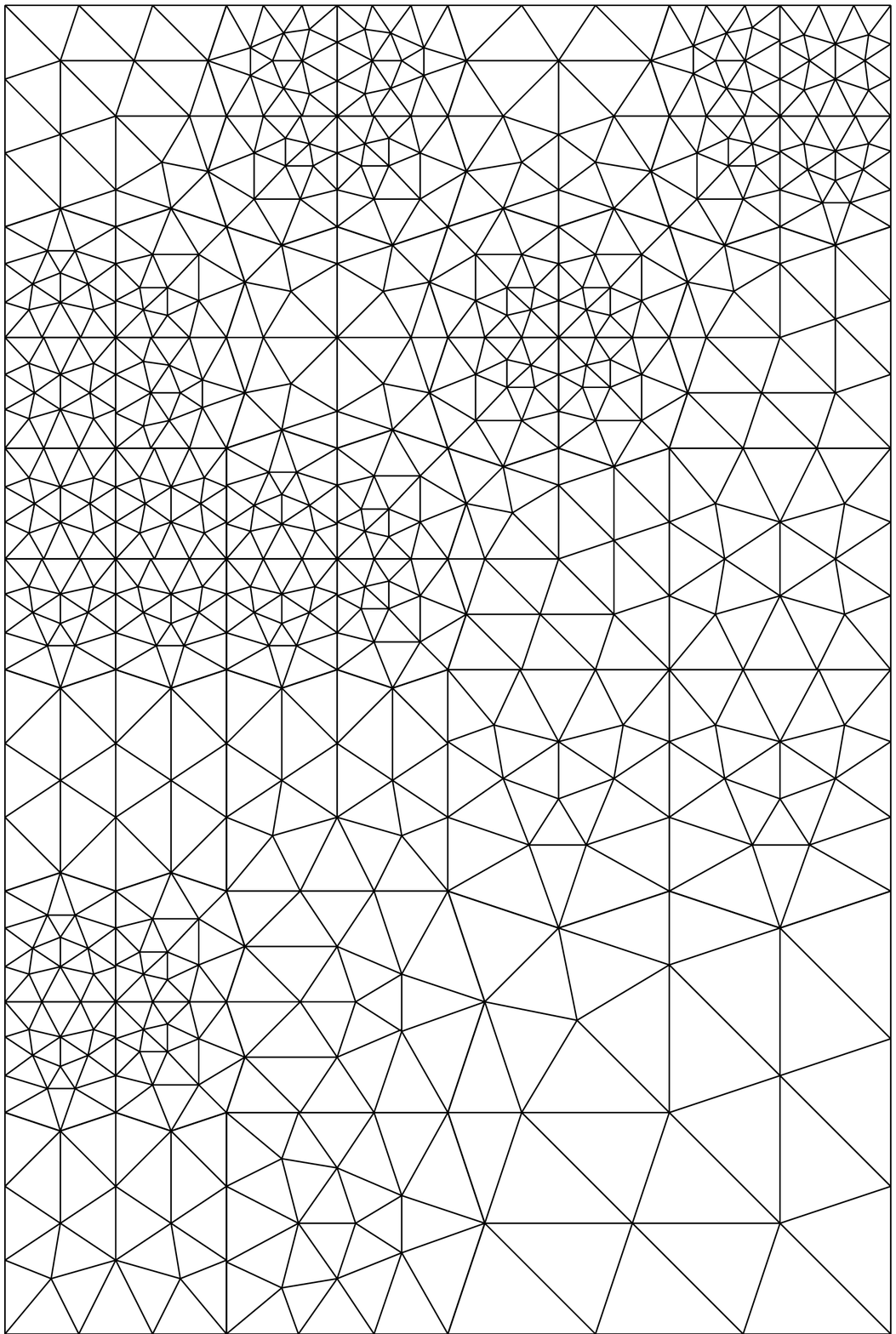,height=2.6in} \\
(a) & (b) 
\end{tabular}
\end{center}
\caption{Acute triangulation gadgets and their use on a point set 
(M. Bern and D. Eppstein~\cite{BernEG94})}
\label{fig:acute_gadgets}
\end{figure}

The acute triangulation problem has been studied for other simple polygons
  as well. Gardner~\cite{Gardner60} asked the question for triangles.
Manheimer proved that seven acute triangles are necessary and sufficient 
  to subdivide a nonobtuse triangle~\cite{Manheimer60}.
Recently, Maehara~\cite{Maehara00} showed that an arbitrary
  quadrilateral can be tiled by $10$ (but perhaps not by any fewer)
  acute triangles.
Gerver~\cite{Gerver84} considered the problem of finding triangulations
  with a stricter upper bound (between $72^\circ$ and $60^\circ$)
  on their angles, and gave necessary conditions for a polygonal
  domain to have such a triangulation.
If we restrict ourselves to two-dimensional point sets, a solution to
  the acute triangulation problem is given by Bern {\em et al.}~\cite{BernEG94}.
Their approach starts with a quadtree, and replaces the squares
  by tiles with protrusions and indentations.
\figr{acute_gadgets} shows sample tiles
  together with an acute triangulation resulting from their algorithm.

Krotov and Krizek~\cite{KrotovK01} studied refinement methods to subdivide 
  a nonobtuse tetrahedral partition into another finer one. 
Unfortunately, they called the resulting triangulations 
  {\em acute type} instead of {\em nonobtuse} even though $90^\circ$ 
  dihedral angles were ubiquitous in them.  
Another related work is by 
  Hangan, Itoh and Zamfirescu~\cite{HanganIZ00, Itoh01} who
  studied acute surface triangulations of certain special shapes
  such as a cube, sphere and icosahedron.

\subsection{Acute and nonobtuse tilings}
\label{sec:tilings}

\begin{figure}[b]
\begin{center}
\begin{tabular}{c c}
\psfig{figure=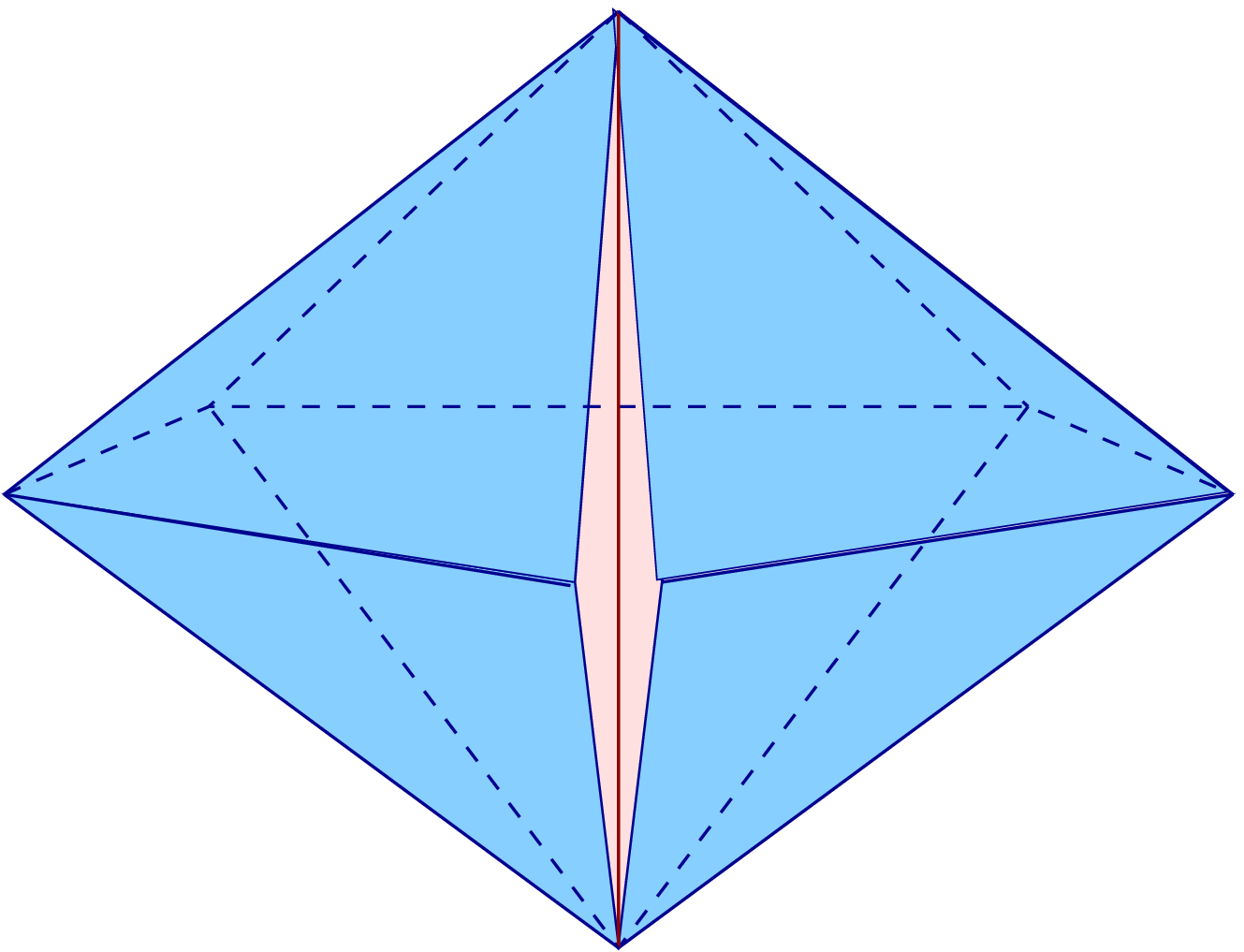,height=1.8in} &
\psfig{figure=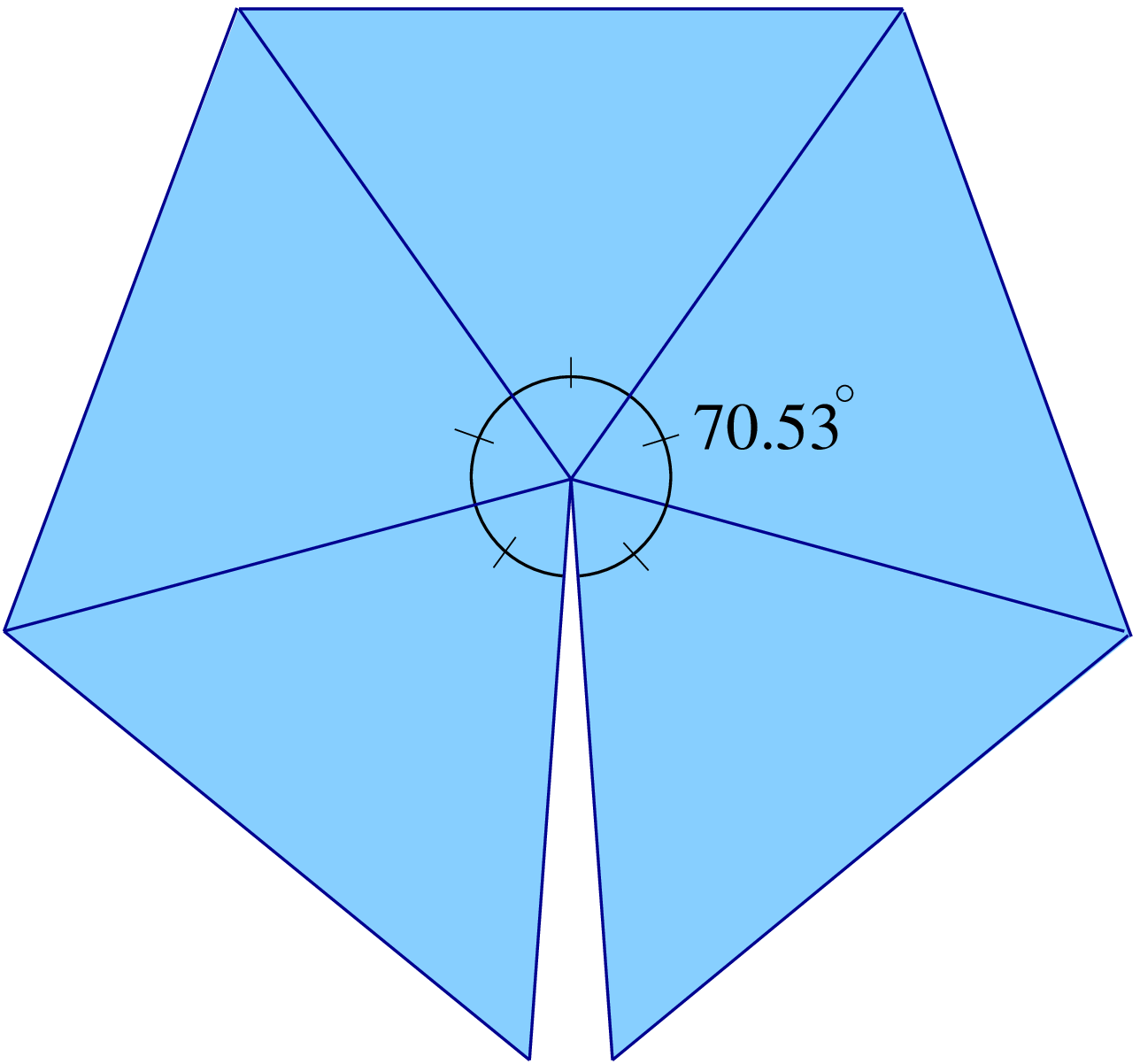,height=1.8in}
\end{tabular}
\end{center}
\caption{The regular tetrahedron does not tile space.}
\label{fig:gap}
\end{figure}

Aristotle claimed that regular tetrahedra could
meet five-to-an-edge to tile space, and this claim
was repeated over the centuries (see~\cite{Senechal81}).
This of course is false, because the dihedral angle of
a regular tetrahedron is not $72^\circ$ but
  $\arccos\tfrac13 \approx 70.53^\circ$.
\figr{gap} shows the small gap left when five tetrahedra are placed around
  an edge.

\begin{figure}[bt]
\begin{tabular}{c c c c}
\psfig{figure=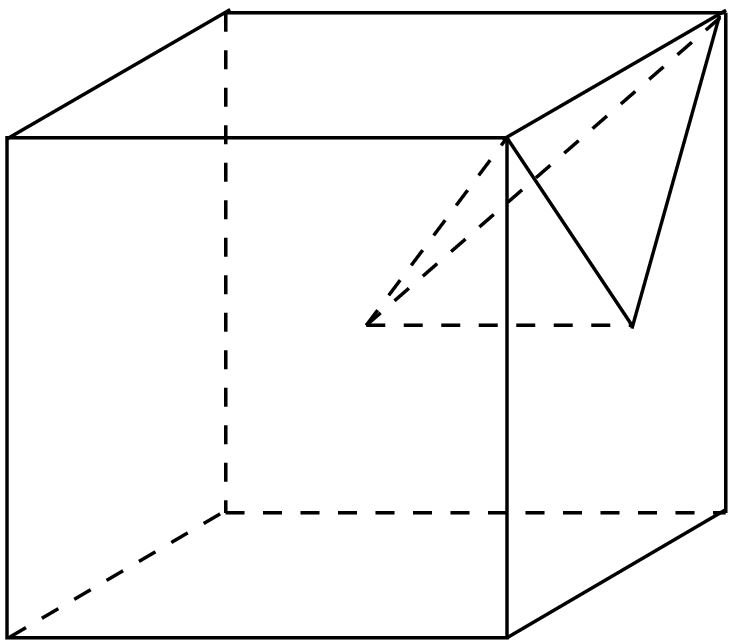,height=1.05in} &
\psfig{figure=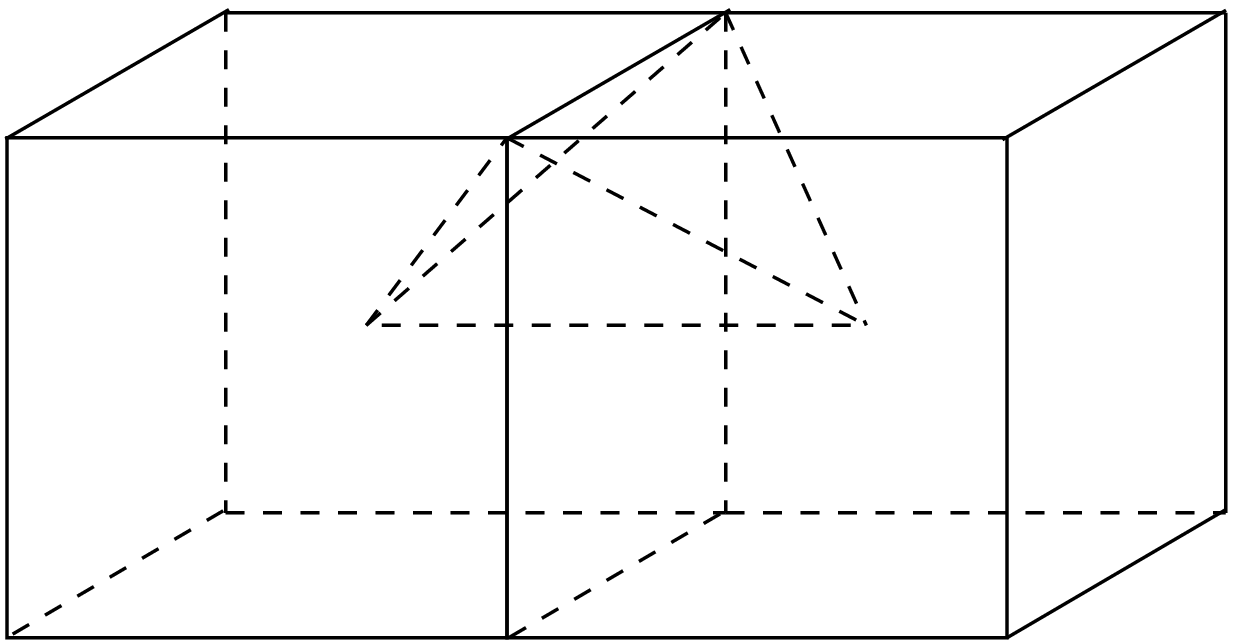,height=1.05in} &
\psfig{figure=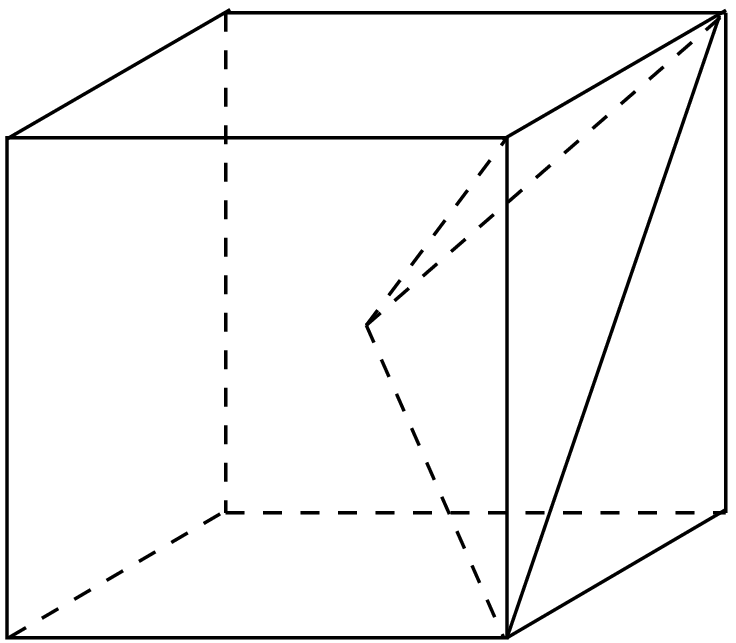,height=1.05in} &
\psfig{figure=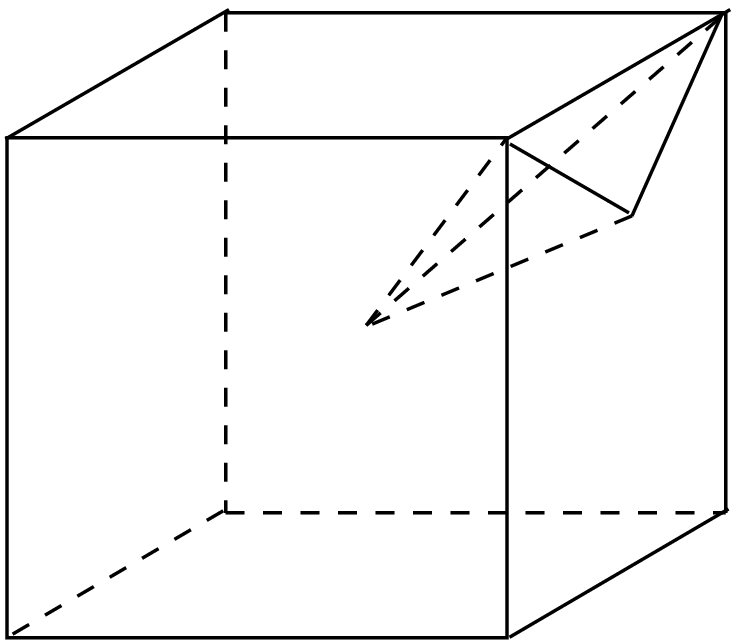,height=1.05in}\\
(a) & (b) & (c) & (d) \\
\end{tabular}
\caption{Sommerville tetrahedra. The first tetrahedron is half of the third. 
The fourth tetrahedron is one fourth of the second.} 
\label{fig:Sommerville}
\end{figure}

\begin{figure}[tb]
\begin{center}
\begin{tabular}{c c}
\psfig{figure=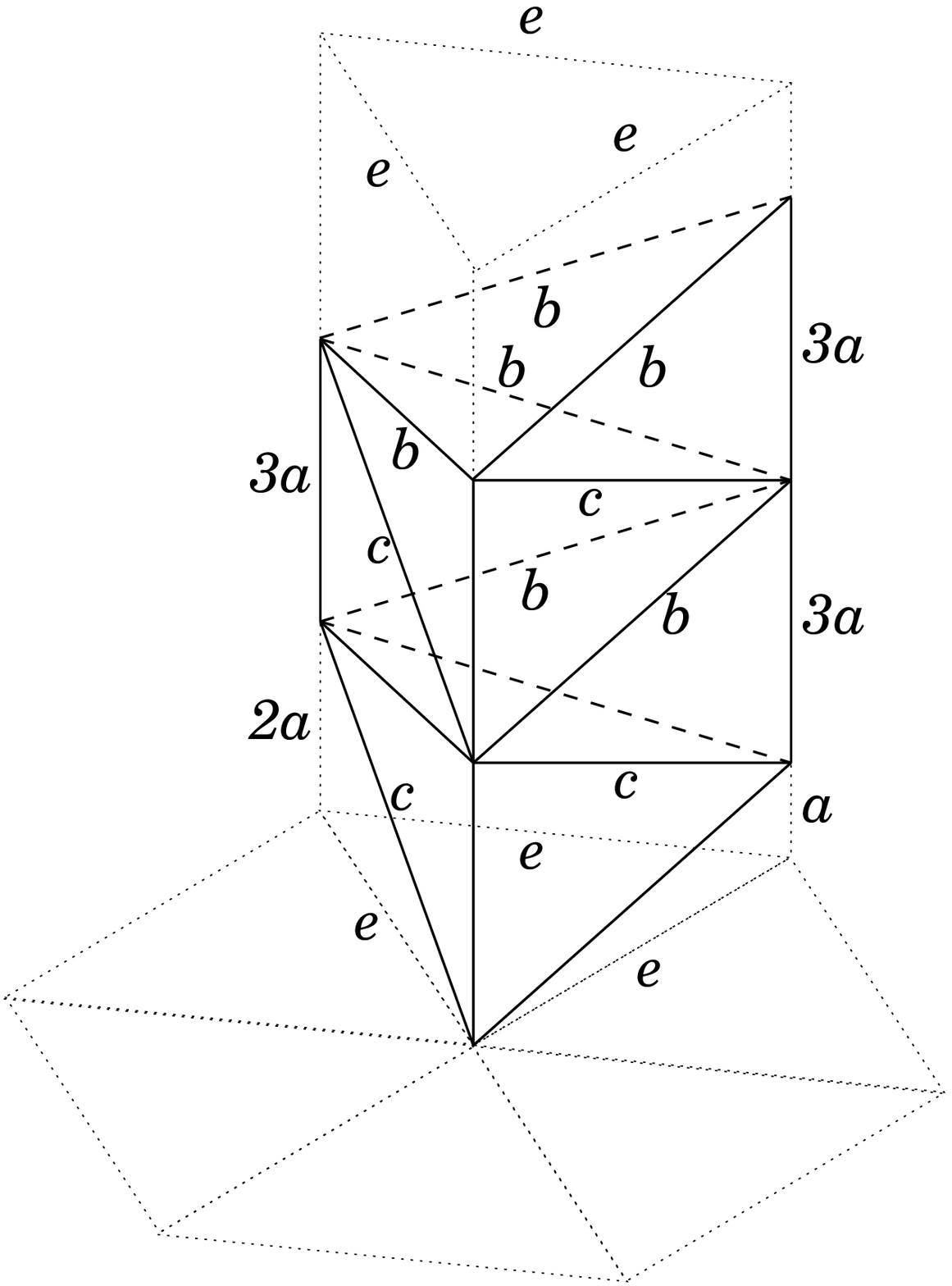,height=3.in} \hspace{.5in}&
\psfig{figure=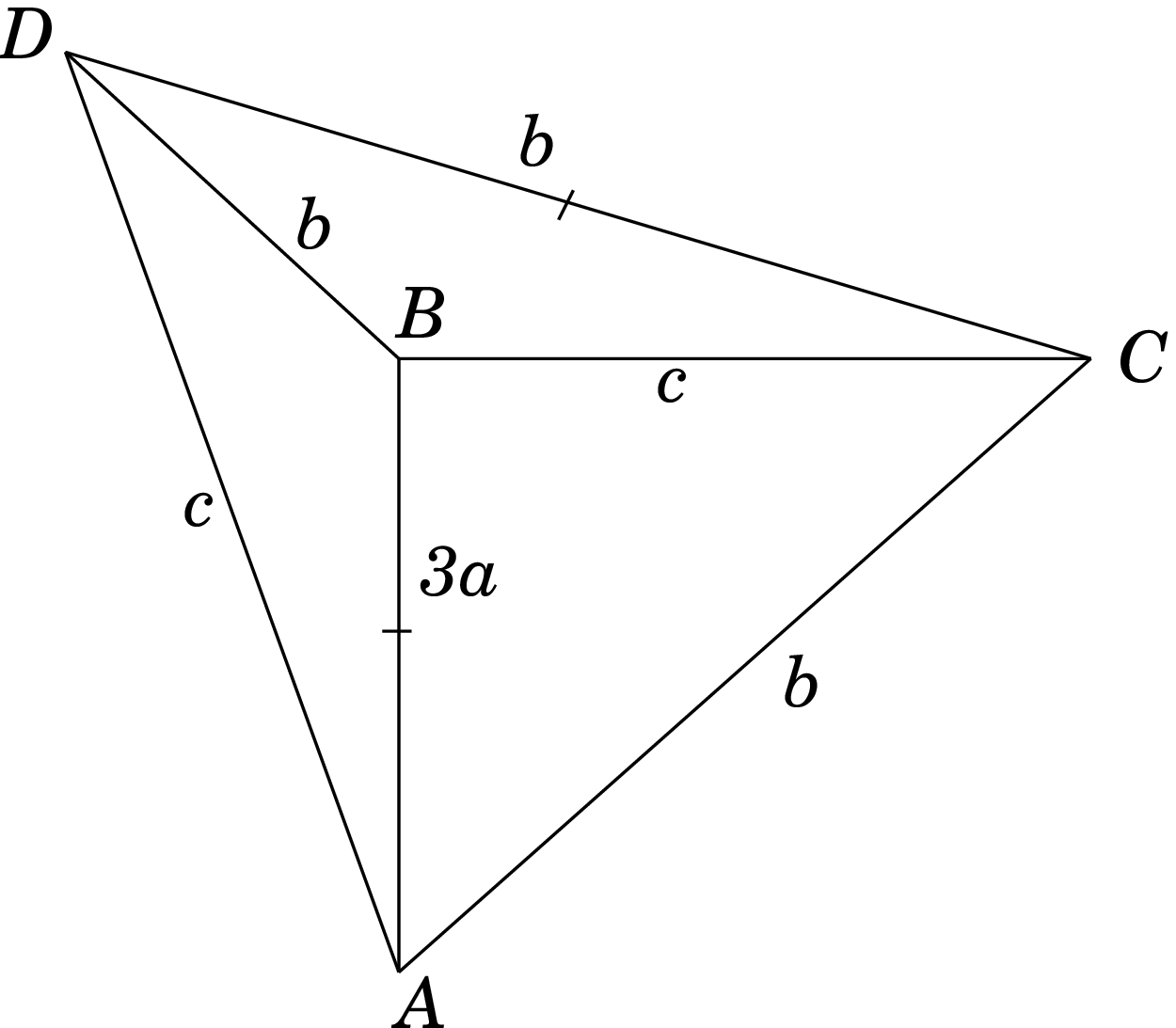,height=2.in}\\
(a) & (b) \\
\end{tabular}
\end{center}
\caption{Family of space tilings}
\label{fig:cons_family}
\end{figure}

There are, however, tetrahedral shapes which can tile space.
Sommerville~\cite{Sommerville23} found four such tetrahedra,
  shown in \figr{Sommerville}.
Four decades later, Davies~\cite{Davies65} and 
  Baumgartner~\cite{Baumgartner68} independently 
  rediscovered three of the Sommerville tetrahedra;
Baumgartner also found a new example. 
Goldberg~\cite{Goldberg74} surveyed the list of all known
  space-tiling tetrahedra, and found three infinite families,
  including the one shown in \figr{cons_family}(a).

The construction of this family is based on a tiling of the 
  plane by equilateral triangles of side-length $e$.
The infinite prism over each triangle is filled with tetrahedra
  whose sides are $3a,b,b,b,c,c$,
  where $b^2 = a^2 + e^2$ and $c^2 = 4a^2 + e^2$,
  as shown in \figr{cons_family}.
Since the ratio $a/e$ is arbitrary, there is a continuous family
  of tetrahedral space-fillers of this type.
Goldberg's two other families can be derived simply by
  cutting these tetrahedra into two conguent pieces,
  either by the plane through $C$, $D$ and the midpoint of $\overline{AB}$,
  or by the plane through $A$, $B$ and the midpoint of $\overline{CD}$.
Whether the list is complete or not is still an open
  problem~\cite{Edelsbrunner01, Senechal81}.
None of the known space-tiling tetrahedra is acute (although
several are nonobtuse).
In fact, the tilings all contain edges of valence four.
(In Goldberg's family, these are the edges of length $c$.)

\begin{figure}[tb]
\begin{center}
\begin{tabular}{c c}
\psfig{figure=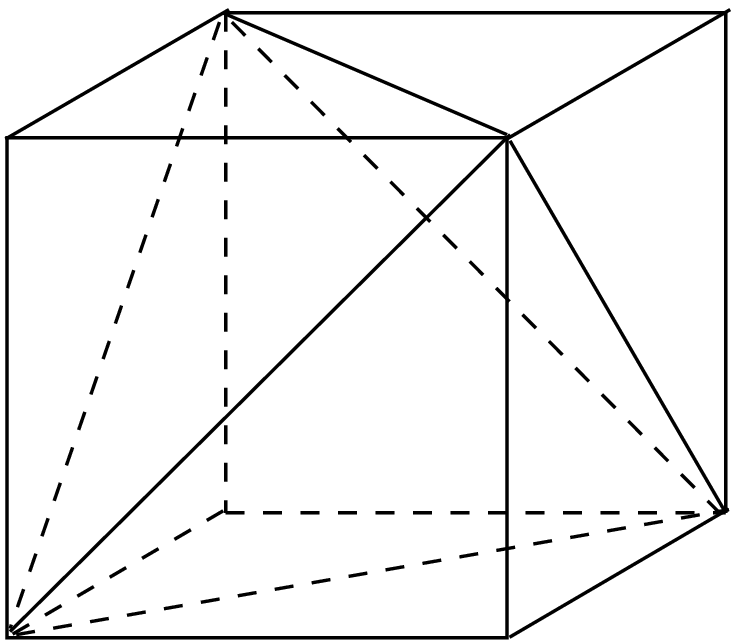,height=1.2in} \hspace{.5in} &
\psfig{figure=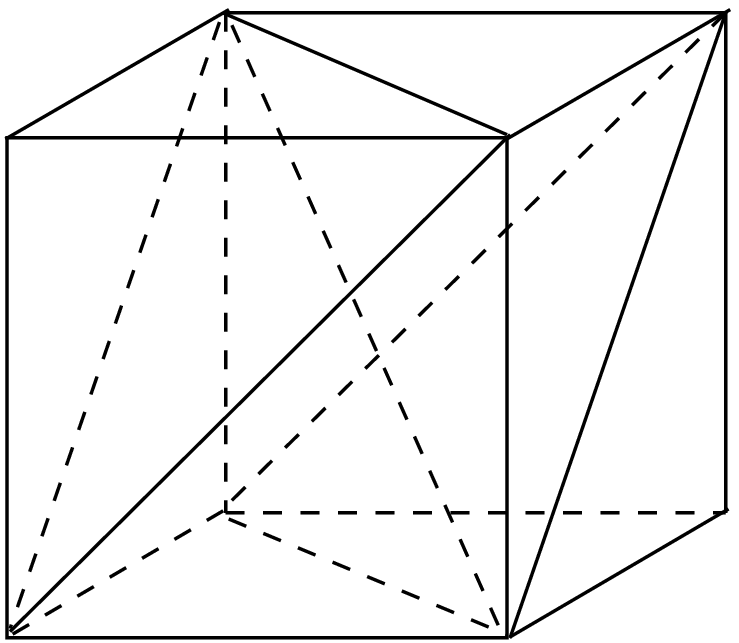,height=1.2in}\\
(a) & (b) \\
\end{tabular}
\caption{Cube subdivided into (a) 5 or (b) 6 tetrahedra.}
\label{fig:cubes}
\end{center}
\end{figure}

Since it seems likely that there is no tiling of space by
congruent acute tetrahedra, we will now consider tilings
with several shapes of tetrahedra.
There are now many more ways to fill space, 
  for instance by subdividing the cube into five or six tetrahedra
  as in \figr{cubes}. 
These tilings also, of course, have $90^\circ$ dihedral angles,
and so are nonobtuse but not acute.

There are many results (like minmax and maxmin angle results)
  known about optimality of Delaunay triangulations in the plane.
But these do not extend to three dimensions, and little is yet
  known about optimum triangulations in space.
Thus it is not surprising that the constructions 
  of acute triangulations do not easily extend from two to three dimensions. 
It is perhaps remarkable that acute triangulations of space
  can be constructed.


\section{Acute Tetrahedra} \label{sec:acuteness}


An acute tetrahedron does not necessarily have high quality in terms of 
  either aspect ratio or radius-edge ratio. 
Low-quality tetrahedra have been classified into nine types~\cite{ChengDEFT99}, 
  and three of these (the spire, splinter and wedge)
  can have all their dihedral angles acute.
However, fortuitously, the tetrahedra in our constructions
  are mostly quite close to regular, and are high-quality
  for use in mesh generation and numerical simulations.



\subsection{Acuteness test}

By definition, a tetrahedron is acute if each of its
  six dihedral angles is less than $90^\circ$.

\begin{figure}[tb]
\begin{center}
\begin{tabular}{c c}
\psfig{figure=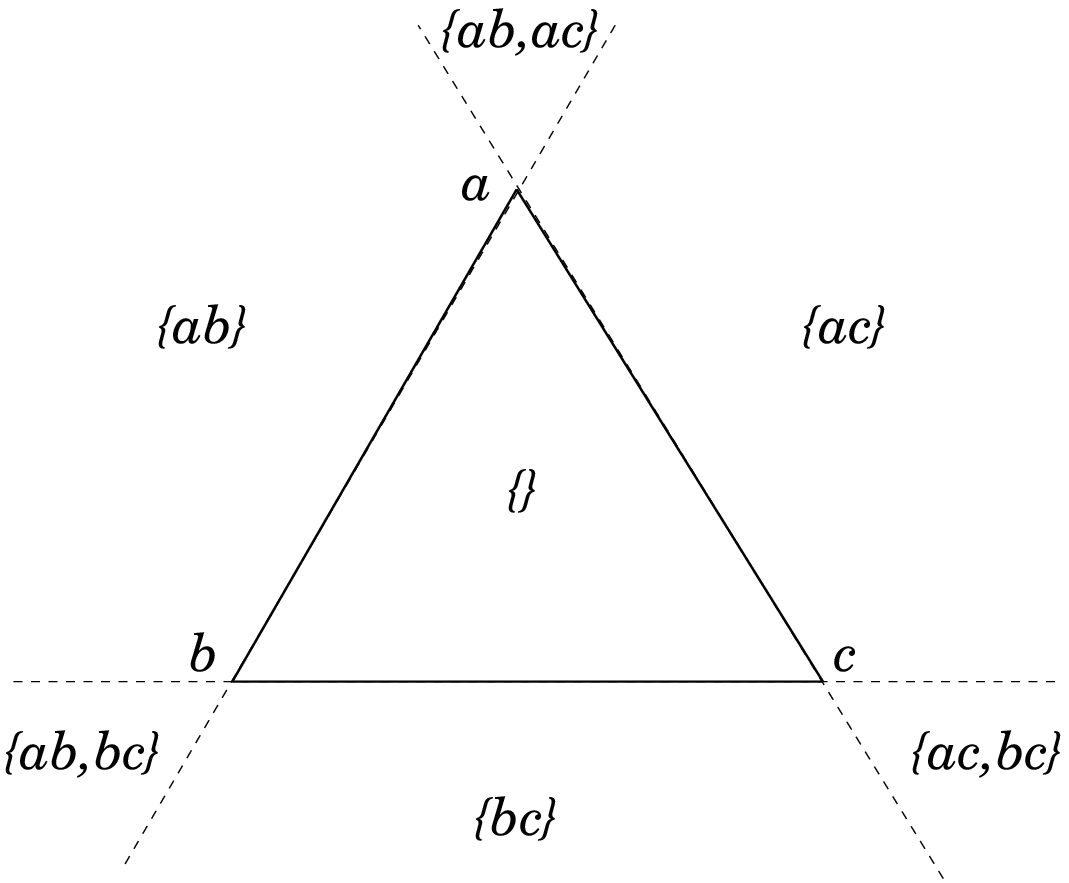,height=1.5in} \hspace{.2in} &
\hspace{.2in}
\psfig{figure=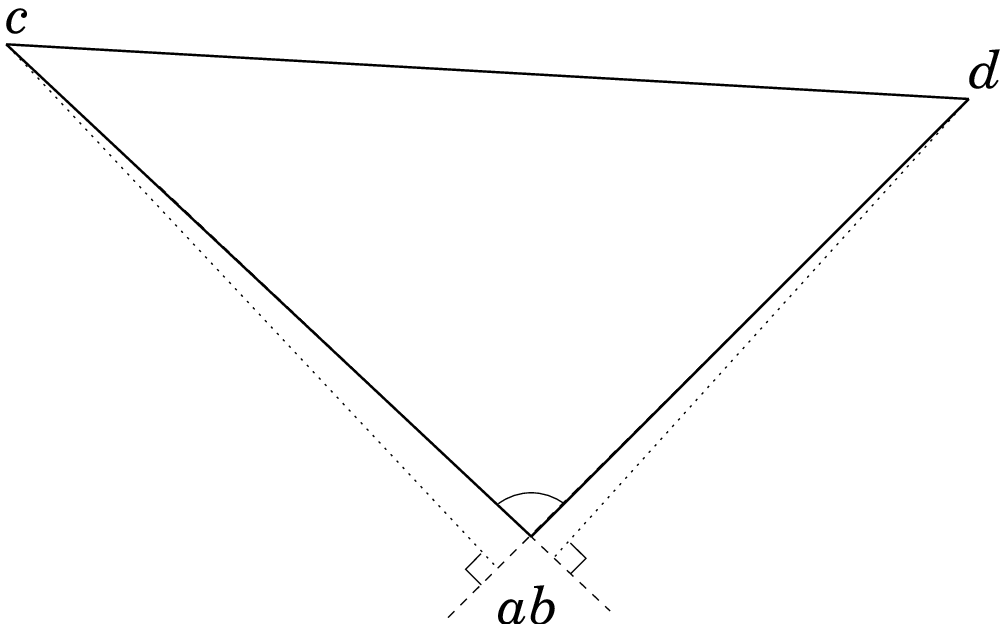,height=1.5in} \hspace{.2in} \\
(a) & (b) \\
\psfig{figure=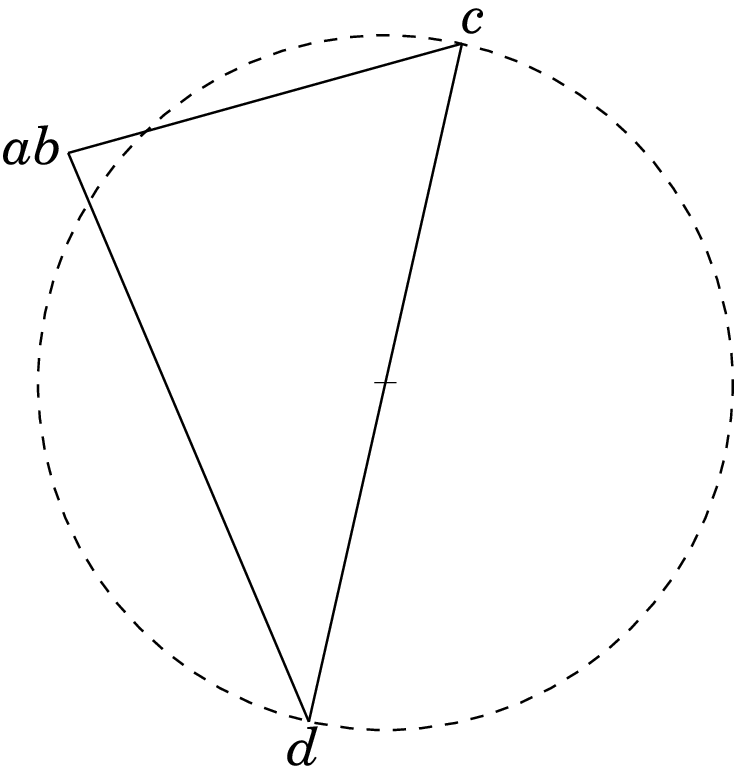,height=1.5in} &
\psfig{figure=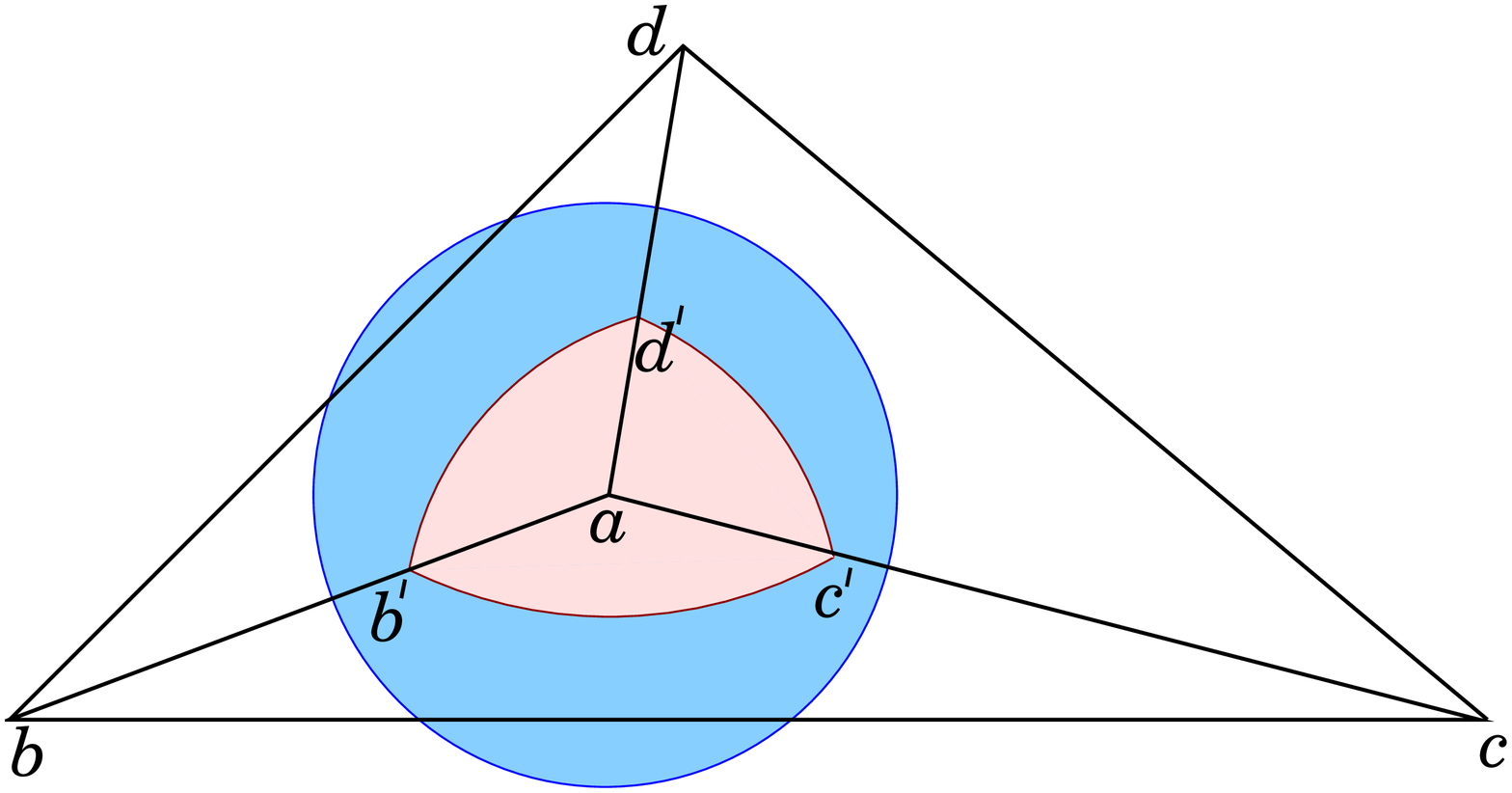,height=1.5in}\\
(c) & (d)
\end{tabular}
\end{center}
\caption{Acuteness tests:
(a) if vertex $d$ projects outside $\triangle abc$,
 the label shows which edges have obtuse dihedral angles;
(b) if the dihedral angle along $ab$ is obtuse, both $c$ and $d$ project
 outside their opposite triangles;
(c) Thales' theorem says that the angle at $ab$ is obtuse exactly when
 it lies outside the circle with diameter $cd$;
(d) if the vertex $d$ projects inside the triangle $abc$ then
 the face angle $\angle bac$ is smaller than the dihedral angle on $ad$.
}
\label{fig:acuteness}
\end{figure}

\begin{lemma}
Consider an edge $ab$ of a tetrahedron $abcd$, and let
$\Pi$ denote projection to a plane normal to $ab$.
The dihedral angle along $ab$ is acute if and only if
$\Pi(a)=\Pi(b)$ lies strictly outside the circle
with diameter $\Pi(c)\Pi(d)$.
\end{lemma}
\begin{proof}
The dihedral angle along $ab$ is by definition the
angle $\angle\Pi(c)\Pi(a)\Pi(d)$; the lemma follows
from standard plane geometry (Thales' theorem, see \figr{acuteness}(c)).
\end{proof}

This lemma can be applied to each of the edges of a tetrahedron.
We now examine some alternate criteria for acuteness.

\begin{lemma}
A tetrahedron is acute if and only if the orthogonal projection of each vertex
onto the plane of the opposite facet lies strictly inside that facet.
An acute tetrahedron has acute facets, but this is not a sufficient condition.
\end{lemma}

\begin{proof}
Suppose the projection of a vertex $d$ is not inside the
opposite triangle $\triangle abc$. 
Then, the dihedral angle along any edge that separates the
 region it projects to from $\triangle abc$ must be nonacute,
 as in \figr{acuteness}(a).
(If the projection is on a triangle edge, then the corresponding
 dihedral angle is exactly $90^\circ$.)
Conversely, if the dihedral angle along edge $ab$ is nonacute,
 then $d$ projects outside $\triangle abc$, as in \figr{acuteness}(b).

We prove the second statement in contrapositive form, while noting
that nonacute sliver tetrahedra can have acute face angles.
Suppose tetrahedron $abcd$ has a nonacute face angle $\angle bac$;
we will show the tetrahedron is nonacute.
If the projection of $d$ onto $\triangle abc$ is not in the interior
we are done by the first part of the lemma.
Otherwise, we claim the dihedral angle along $ad$ is larger than
  $\angle{bac}$ and thus is nonacute.
To check the claim, remember
the spherical dual law of cosines (see \cite{Thurston97}):
$$\cos d' = -\cos b' \cos c' + \sin b' \sin c' \cos \angle{bac}$$
  where $b'$, $c'$ and $d'$ are the dihedral angles along the edges 
  $ab$, $ac$ and $ad$, respectively.  (See \figr{acuteness}(d).)
Assuming $b', c' < \tfrac \pi 2$, this gives
$\cos d' <  \cos \angle{bac}$ as desired.
\end{proof}

\subsection{Acuteness of Delaunay Triangulations}

Given a set of vertices, the Delaunay triangulation is optimal in many ways.
However, a Delaunay triangulation in any dimension can have obtuse angles.
In this section, we investigate the converse, whether an acute
triangulation is necessarily the Delaunay triangulation for its vertices.
The answer is positive in the plane, but negative in three-space.

\begin{lemma}
Any acute two-dimensional triangulation $\cal{T}$ is Delaunay.
\end{lemma}

\begin{figure}[htb] \begin{center}
\psfig{figure=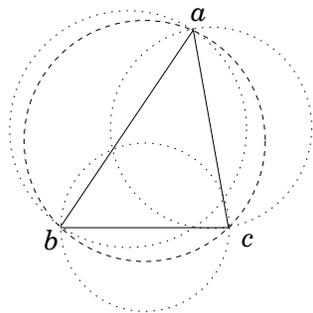,height=1.6in}
\caption{Acute triangles in the plane are Delaunay.  An alternate
proof uses the fact that the circumcircle is contained in the union
of the three diametral circles around the edges.}
\label{fig:acuteDel2d}
\end{center} \end{figure}

\begin{proof}
Since $\cal{T}$ is acute, the diametral circle of each edge
is empty of other vertices.  By definition, this means the edge
is in the Gabriel graph of the vertex set, which is a subgraph
of the Delaunay triangulation.
But since the edges of $\cal{T}$ form a triangulation, it must be the
entire Delaunay triangulation.  See also \figr{acuteDel2d}.
\end{proof}

\begin{corollary}
If an acute triangulation of a two-dimensional vertex set exists
  then it is unique. 
\qed
\end{corollary}

\begin{lemma}
There is an acute triangulation $\cal{T}$ in three dimensions
which is not Delaunay.
\end{lemma}
\begin{figure}[htb] \begin{center}
\psfig{figure=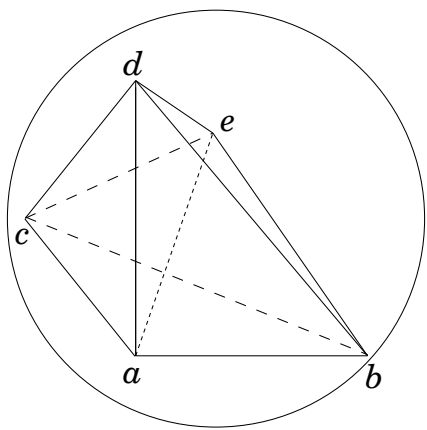,height=2.2in}
\caption{An example of an acute triangulation in space which is not Delaunay.}
\label{fig:acuteDel3d}
\end{center} \end{figure}

\begin{proof}
Consider a ``cube corner'' tetrahedron, and glue it to a copy
of itself across the equilateral face.  Then move the two corner
vertices away from each other a tiny amount to make the tetrahedra acute.
In coordinates, take a suitable small $\epsilon>0$, and let 
$a=(-\epsilon,-\epsilon,-\epsilon)$, 
$b=(1,0,0)$, $c=(0,1,0)$, $d=(0,0,1)$, and 
$e=(2/3+\epsilon, 2/3+\epsilon, 2/3+\epsilon)$.
The two tetrahedra $abcd$ and $bcde$ are acute, but
the Delaunay triangulation of these five points 
consists of three tetrahedra: $abce$, $acde$, and $abde$,
as in \figr{acuteDel3d},
because $e$ is inside the circumsphere of $abcd$.
Note that the three Delaunay tetrahedra are obtuse,
having $120^\circ$ dihedral angles along edge $ae$.
The acute triangulation we started with is obtained by
performing a 3-to-2 flip on the Delaunay triangulation.
\end{proof}

\section{Constructions for Acute Tilings} \label{sec:constructions}

\subsection{TCP triangulations} 

Our first set of acute triangulations basically come from
the crystallography literature.  Chemists studying alloys
of two transition metals have often found that since the
two types of atoms are similar (but slightly different)
in size, the Delaunay triangulation of their positions
is built of nearly regular tetrahedra.  These TCP
(tetrahedrally close packed) structures were first described
by Frank and Kasper~\cite{FraKas58,FraKas59} and have been
studied extensively by the Shoemakers~\cite{ShoSho86} among others.

A combinatorial definition of the TCP class was given by
Sullivan~\cite{Sullivan99}:
A triangulation is called TCP if every
edge has valence 5 or 6, and no triangle has two 6-valent edges.
This definition includes all the chemically known TCP structures,
but also allows some new structures~\cite{Sullivan00} not yet
seen in nature.

\begin{figure}[htb]
\begin{center}
\begin{tabular}{c c c c}
\psfig{figure=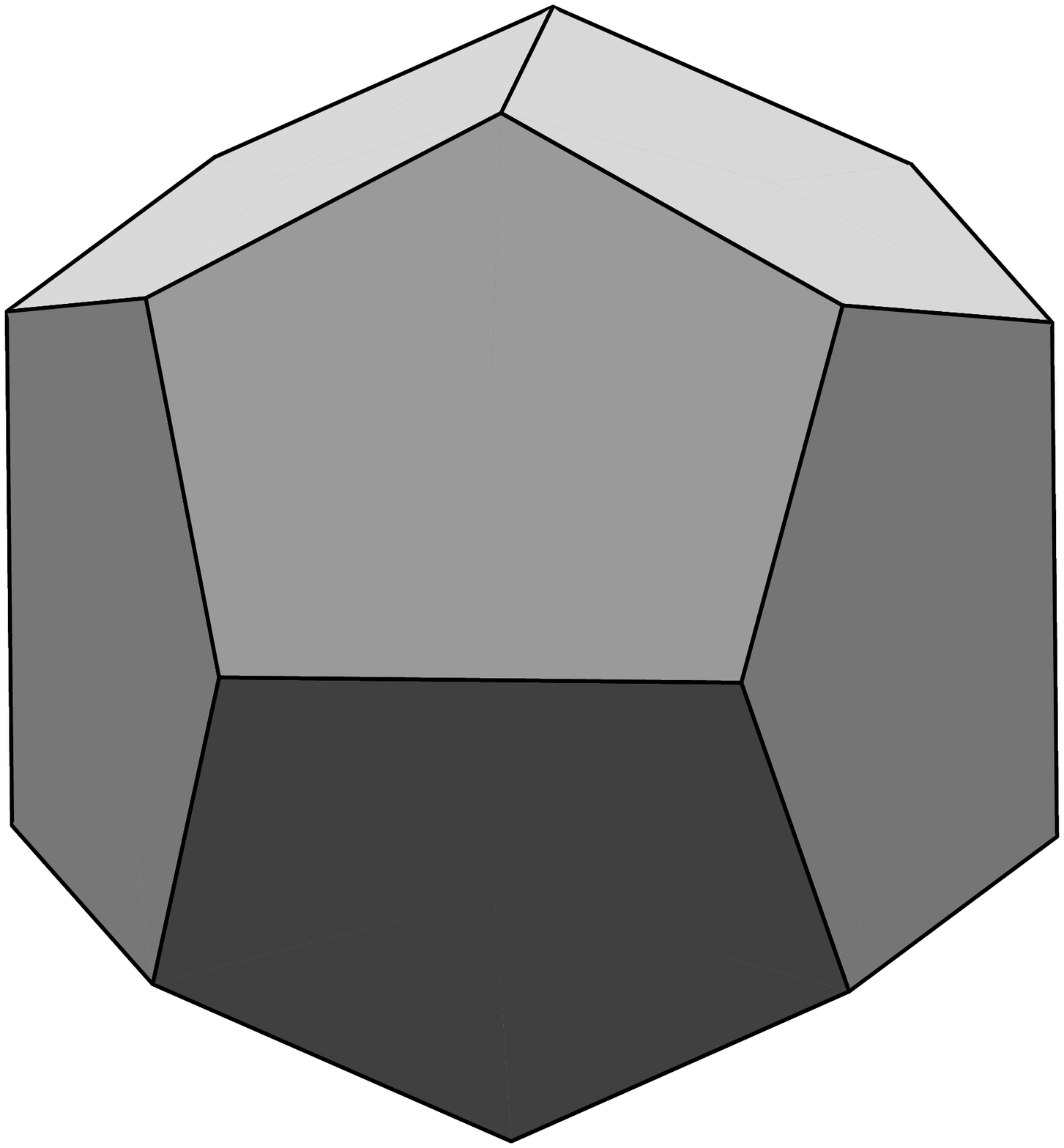,height=.9in}& 
\psfig{figure=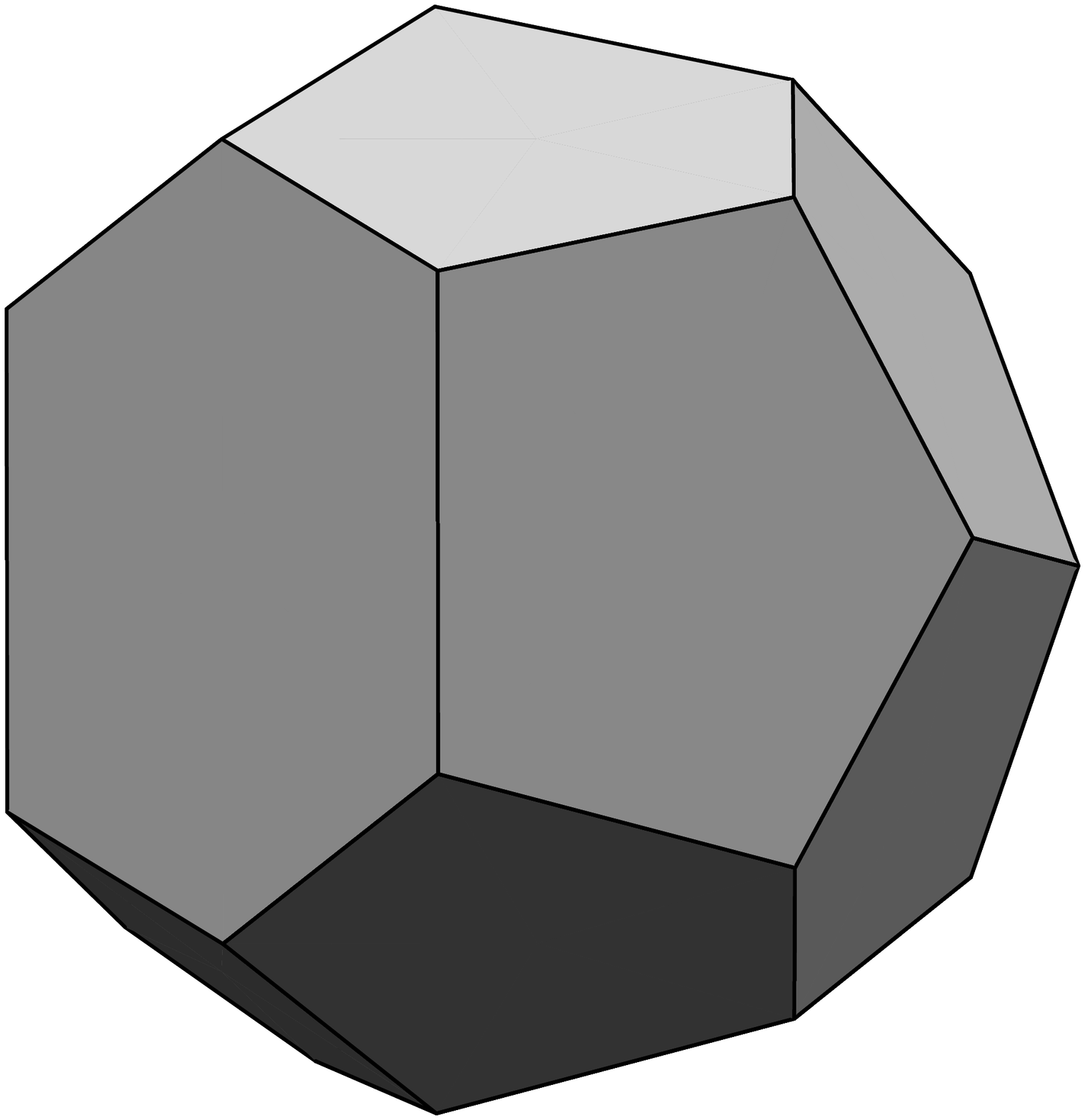,height=.9in}& 
\psfig{figure=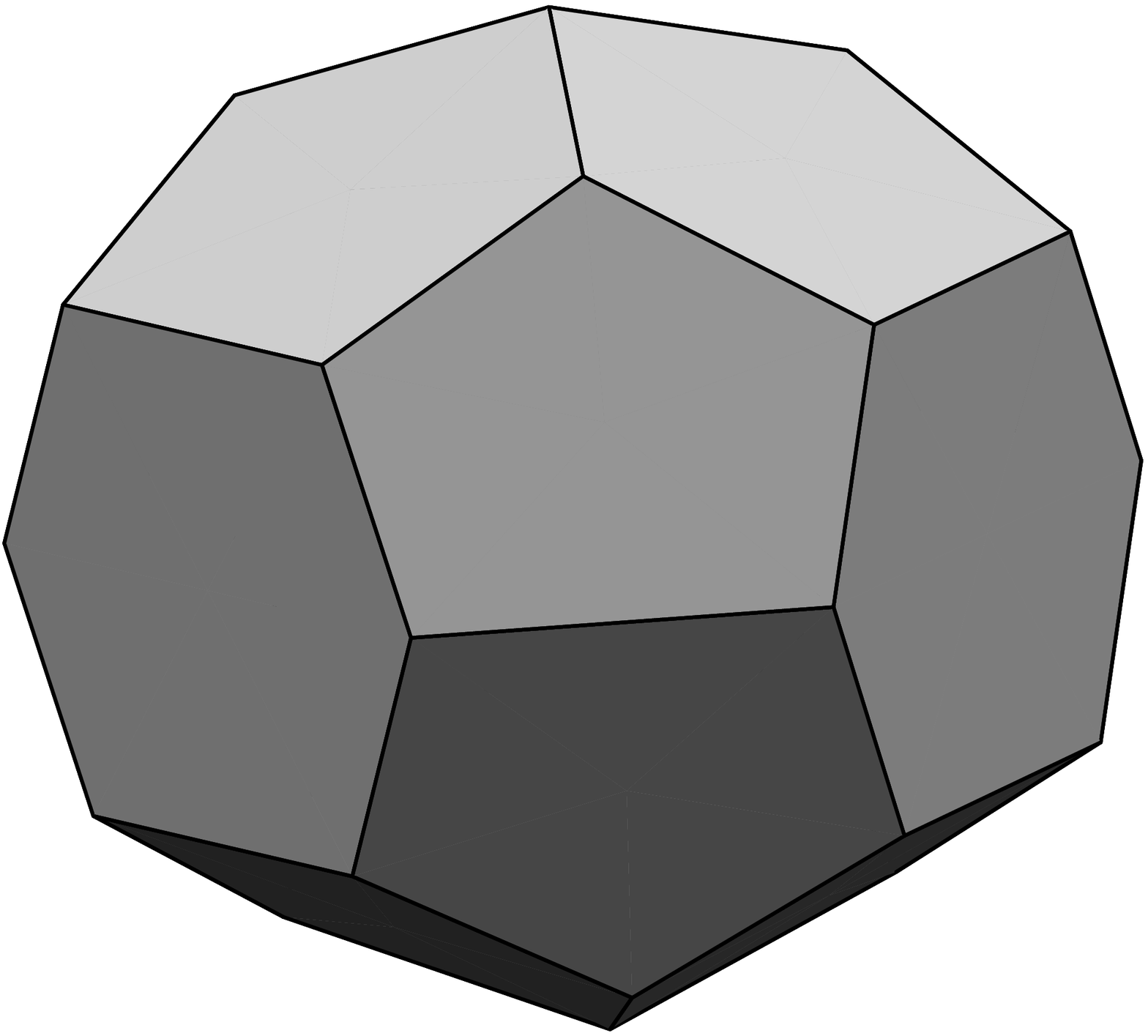,height=.9in}& 
\psfig{figure=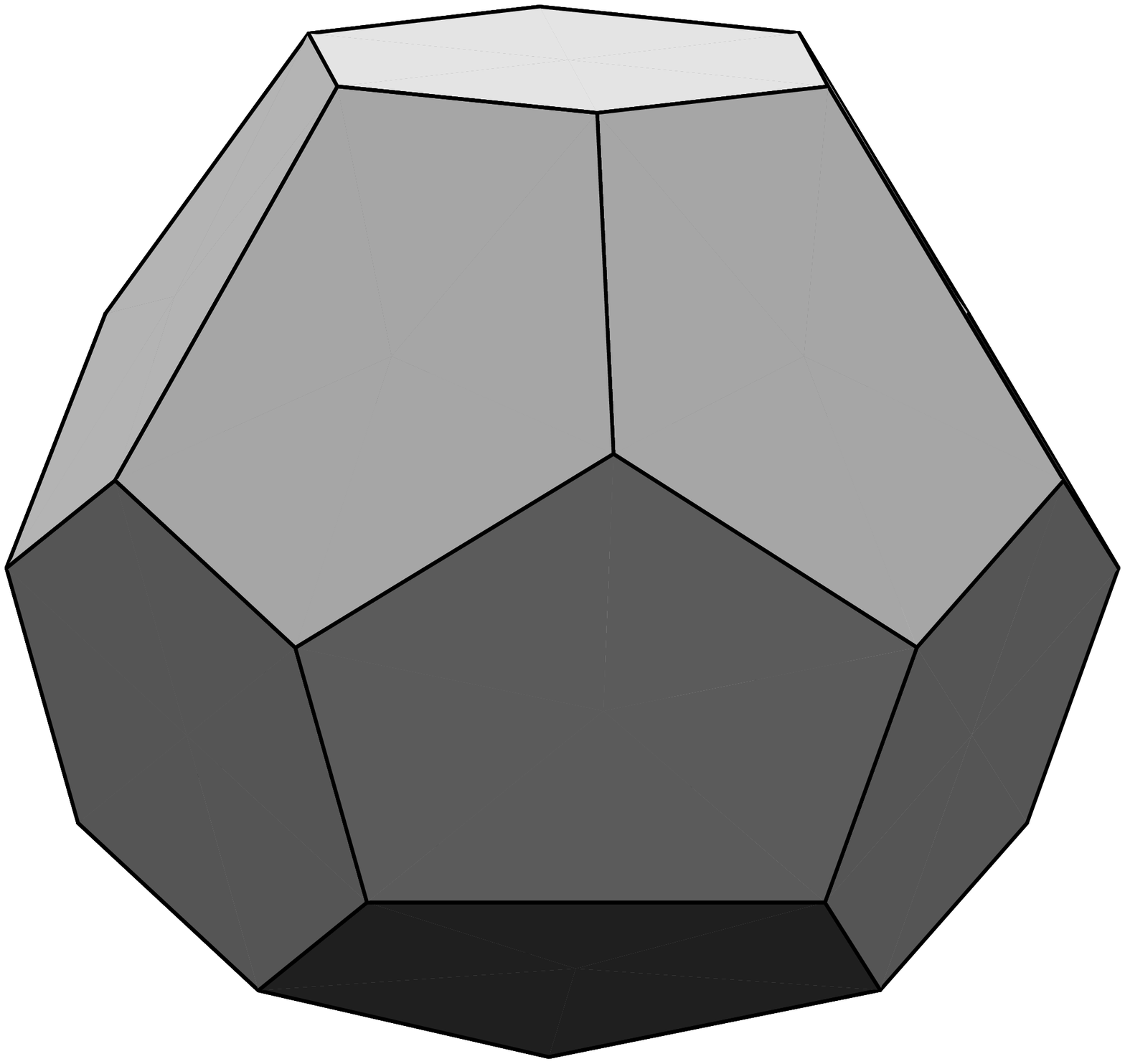,height=.9in}\\
(a) & (b) & (c) & (d) \\
\end{tabular}
\caption{Foam cells with pentagonal and hexagonal faces}
\label{fig:foam_cells}
\end{center}
\end{figure}

It is not hard to check that the definition allows exactly
four types of vertex star in a TCP triangulation.  Dually,
the voronoi cell around any vertex has one of the four combinatorial
types shown in \figr{foam_cells}: these are the polyhedra with pentagonal
and hexagonal faces but no adjacent hexagons.
(It is interesting that these dual structures are
seen in some other crystal structures:
in some zeolites, silicon dioxide outlines the voronoi edges,
while in clathrates, water cages along the voronoi skeleton trap
large gas molecules.)

\begin{figure}[tb] \begin{center}
\begin{tabular}{c c c}
\psfig{figure=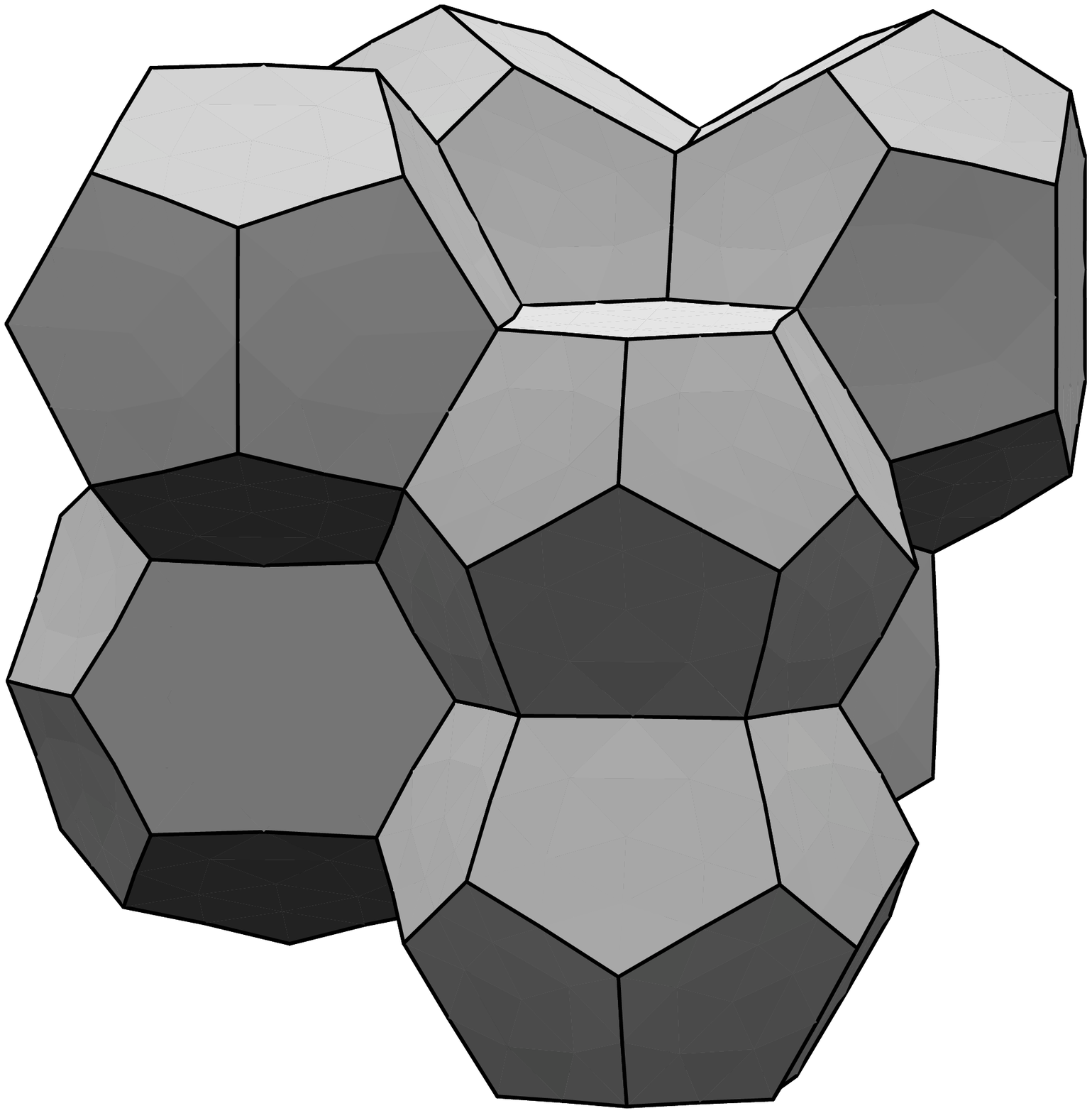,height=1.4in}& 
\psfig{figure=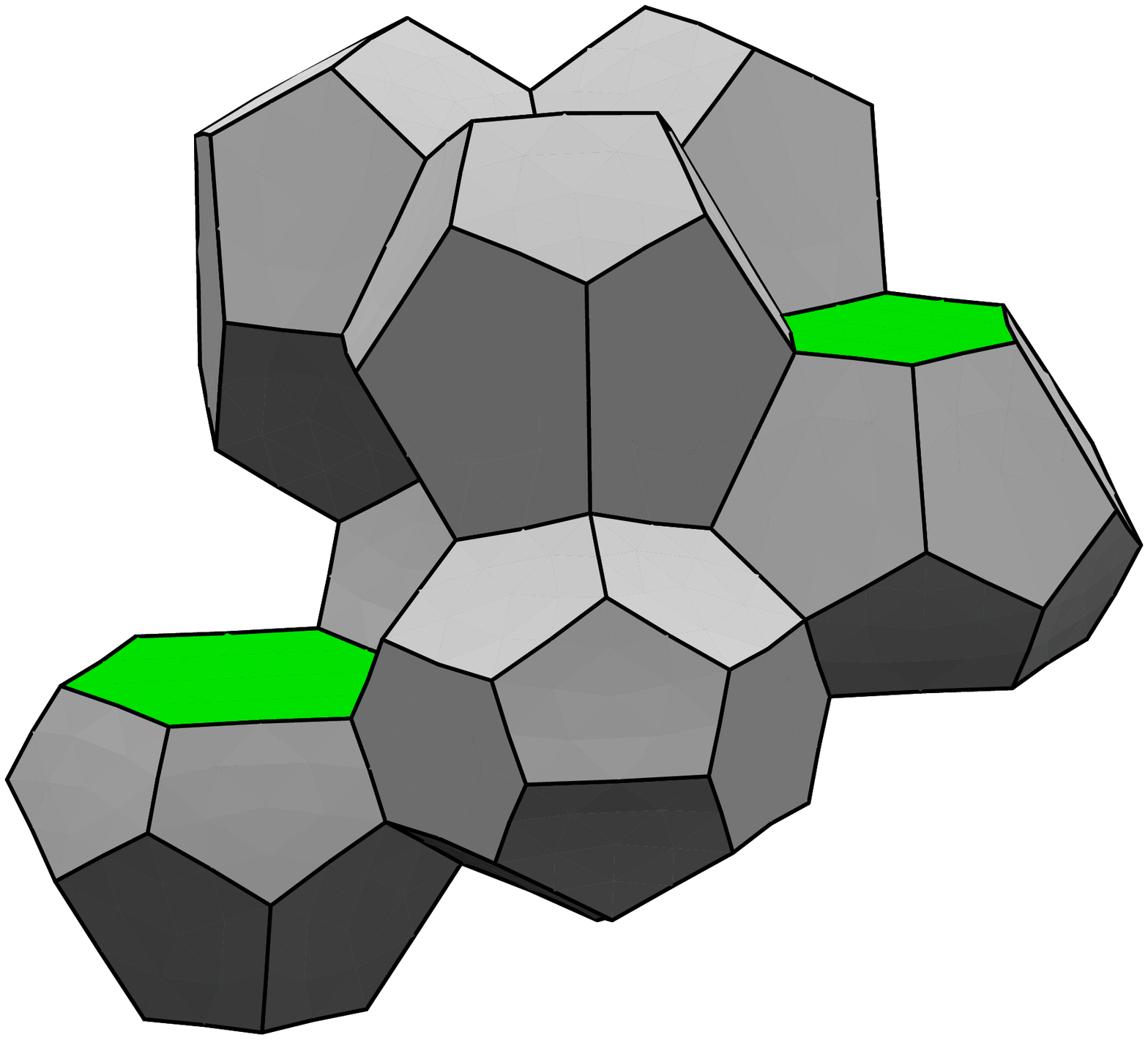,height=1.4in}& 
\psfig{figure=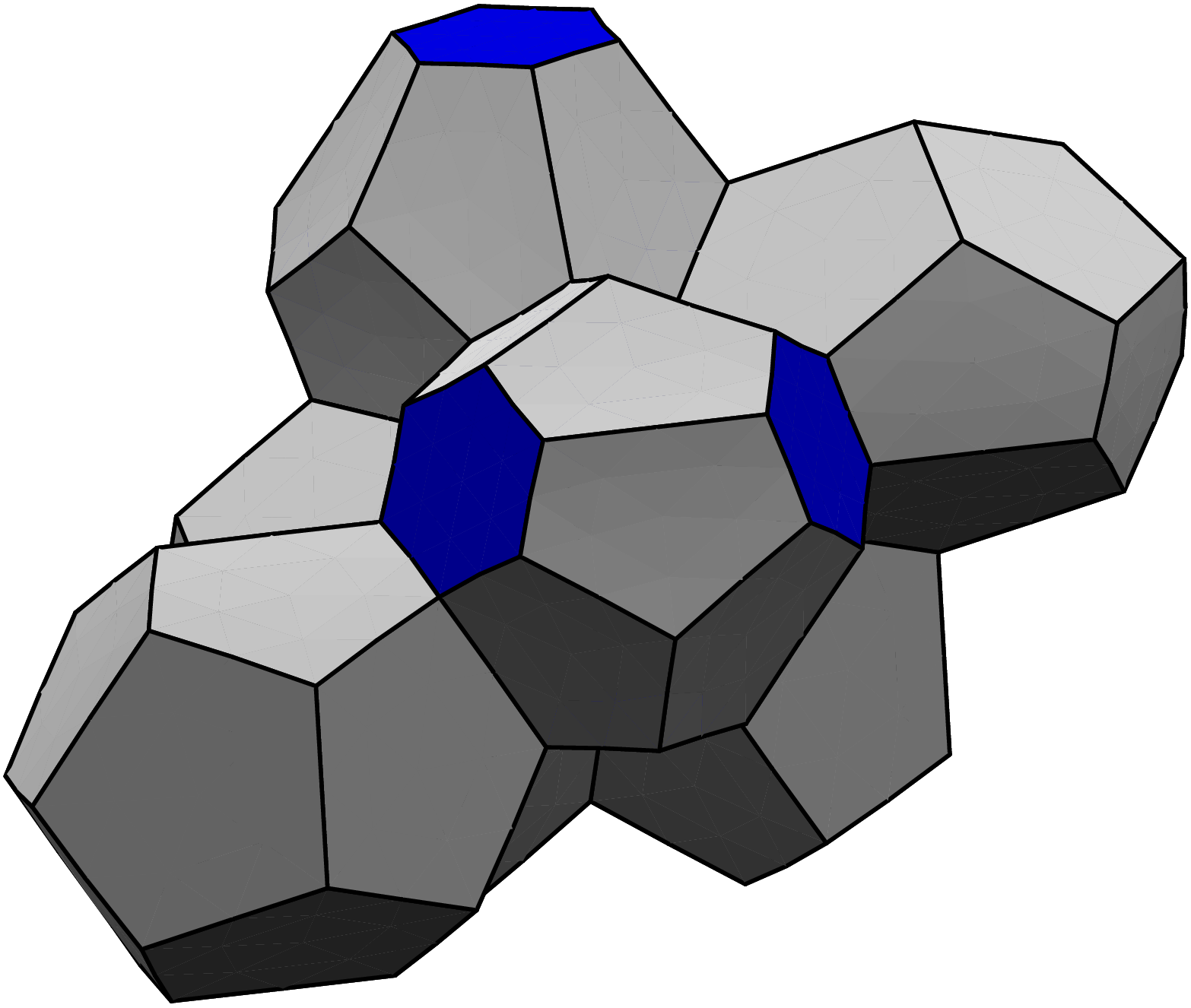,height=1.4in}\\
(a) & (b) & (c)\\
\end{tabular}
\caption{The voronoi cells for the three basic TCP structures, A15, Z, and C15.}
\label{fig:foams}
\end{center} \end{figure}

All known TCP structures can be viewed as convex combinations of
the three basic ones (called A15, Z and C15) shown in \figr{foams}.
There are many ways to understand these structures~\cite{Sullivan99}.
To construct A15, we can start with a BCC lattice.  Its Delaunay
triangulation is one of the Sommerville tilings discussed
above; since the edges have even valence the tetrahedra
can be colored alternately black and white.  If we take the BCC
lattice together with the circumcenters of all black tetrahedra,
we have the vertices of A15: their Delaunay tetrahedra are all now
nearly regular.  Similarly, the C15 structure arises from the diamond
lattice by adding selected circumcenters, and the Z structure can
be obtained similarly starting with hexagonal prisms.

\begin{figure}[htb]
\begin{center}
\begin{minipage}{.25\textwidth}
\begin{tabular}{c}
\psfig{figure=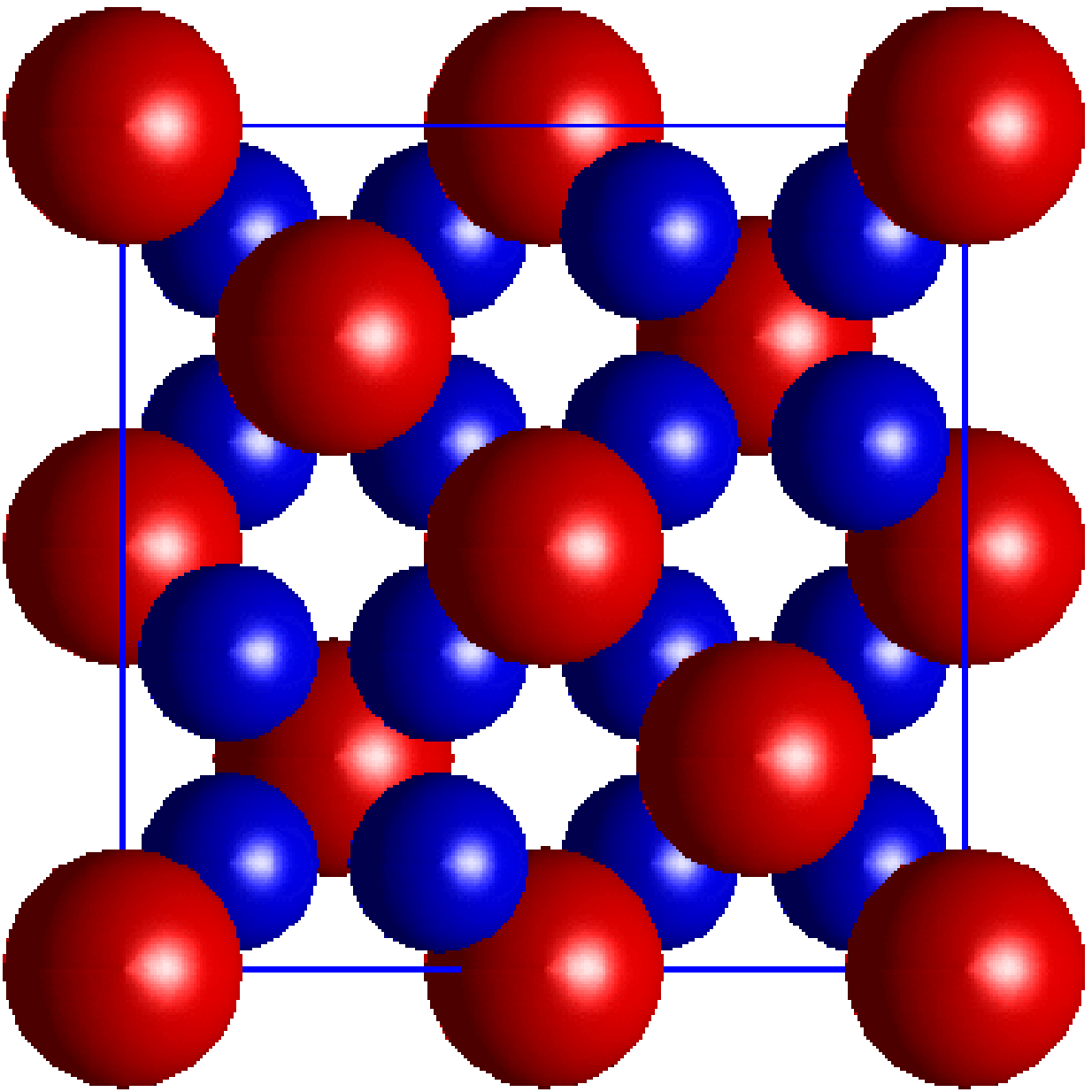,width=.9\textwidth}\\ (a)
\end{tabular}
\end{minipage}
\hspace{.2in}
\begin{minipage}{.25\textwidth}
\begin{tabular}{c}
\psfig{figure=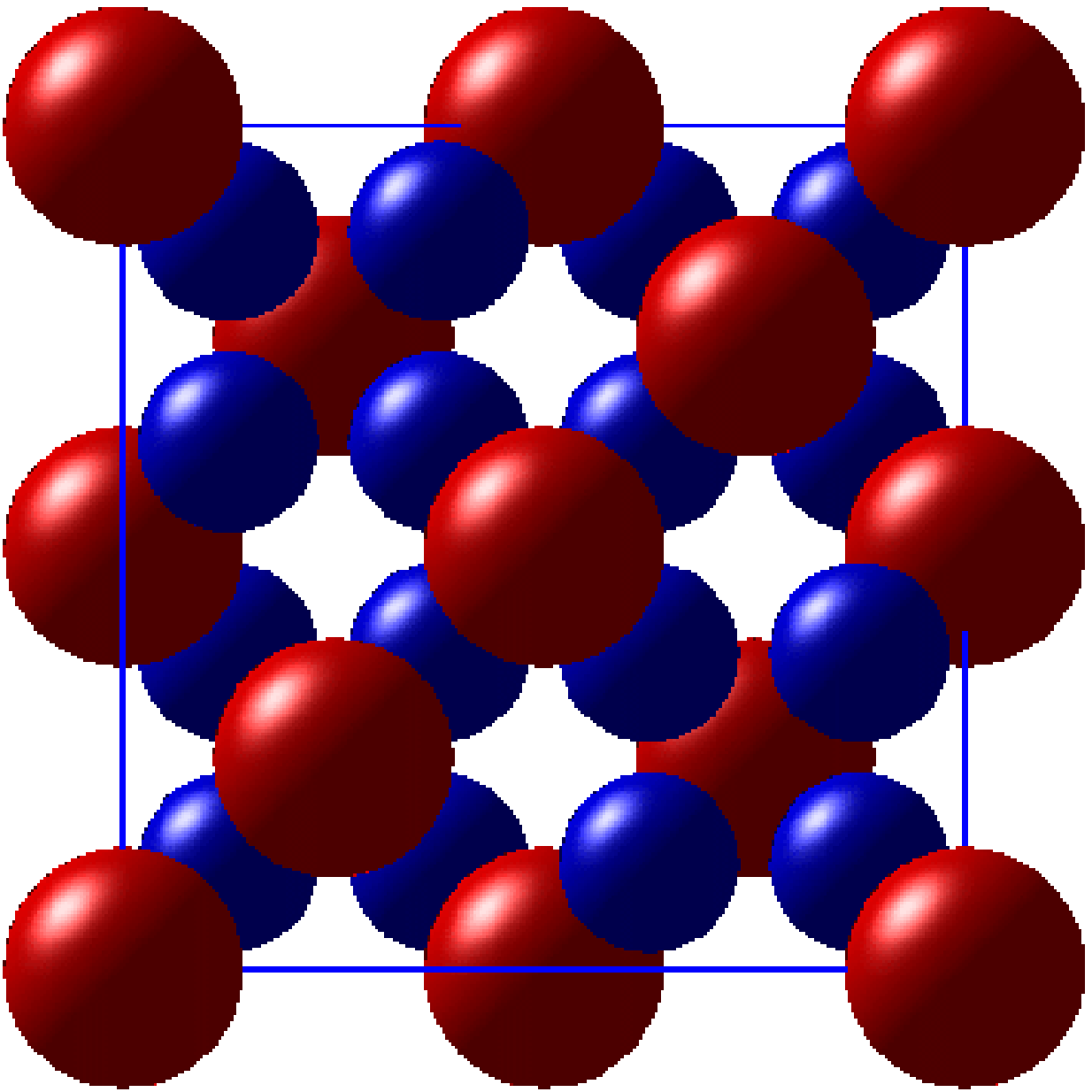,width=.9\textwidth}\\ (b)
\end{tabular}
\end{minipage}
\hspace{.2in}
\begin{minipage}{.25\textwidth}
\begin{tabular}{c}
\psfig{figure=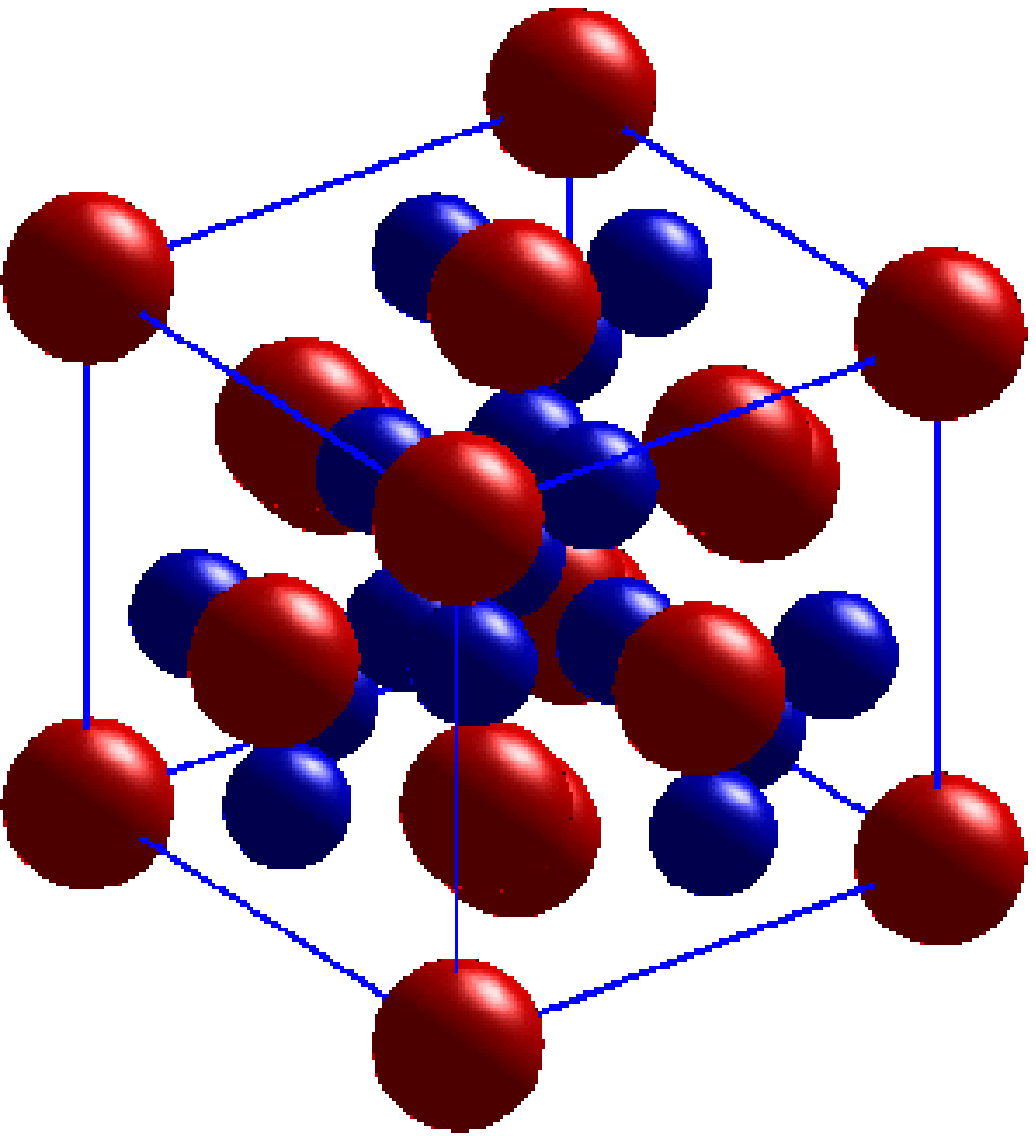,width=.9\textwidth} \\ (c)
\end{tabular}
\end{minipage}
\end{center}
\caption{The vertices of the C15 triangulation
are at the centers of these balls.}
\label{fig:C15}
\end{figure}

The C15 structure (also known as the cubic Friauf--Laves phase)
is shown in \figr{C15}, where the red spheres are centered on a diamond
lattice (FCC together with a certain translate) and the blue
spheres are at selected circumcenters.

In any triangulation of space, the average dihedral angle multiplied
by the average edge valence is exactly $360^\circ$.  If a tiling
could be made of regular tetrahedra, the average edge valence would
thus be $n_0 := 360^\circ/\arccos(\tfrac13) \approx 5.1043$.
But by symmetry, the regular tetrahedron is a critical point for
average dihedral angle, so any tiling made of nearly regular tetrahedra
should have average valence quite close to $n_0$.  Indeed, all known
TCP structures have average valence between $5\tfrac1{10}$
and $\le 5\tfrac19$, the values for C15 and A15.

\begin{figure}[tb] \begin{center}
\begin{tabular}{c c}
\psfig{figure=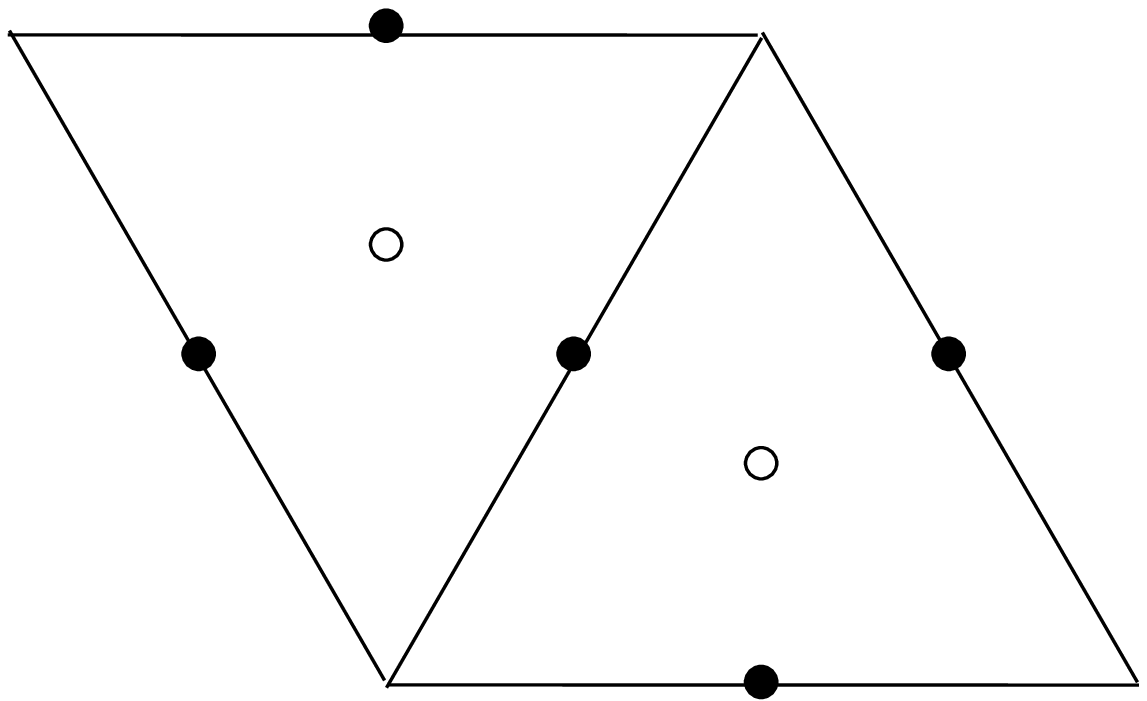,height=.7in}& 
\psfig{figure=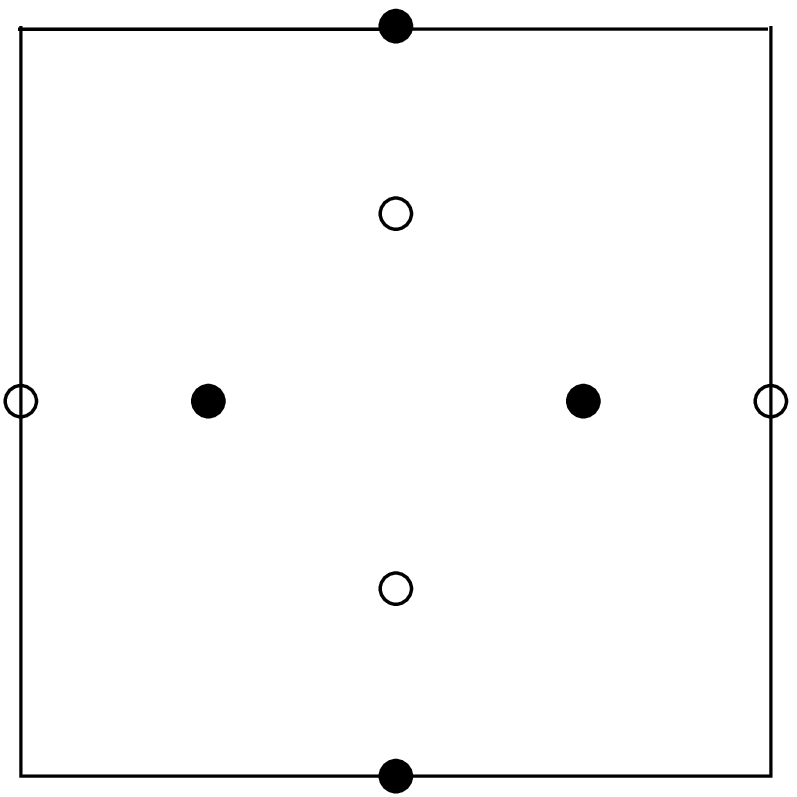,height=.8in}\\
(a) & (b) \vspace{.1in} \\
\psfig{figure=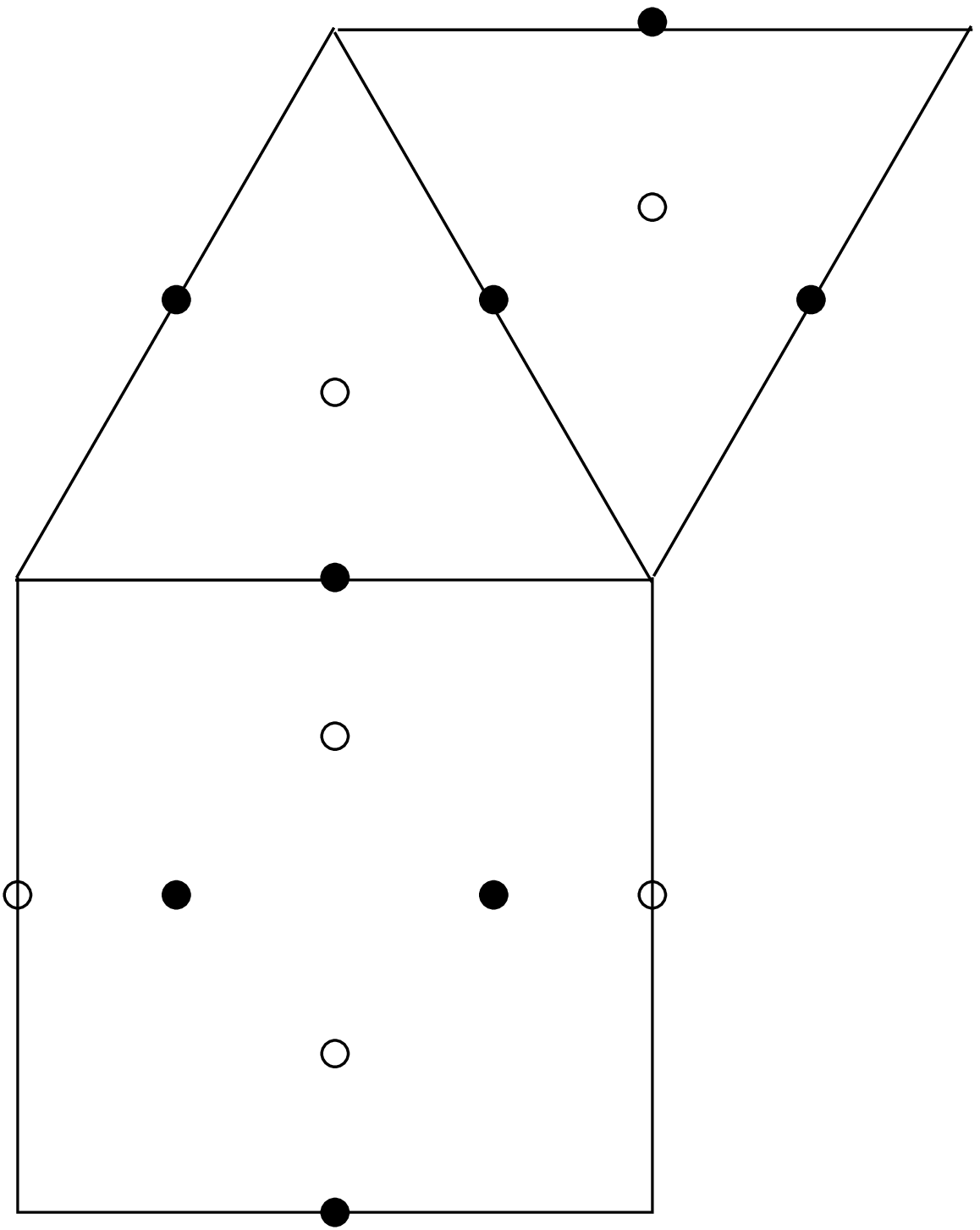,height=1.4in} & \hspace{-.2in}
\psfig{figure=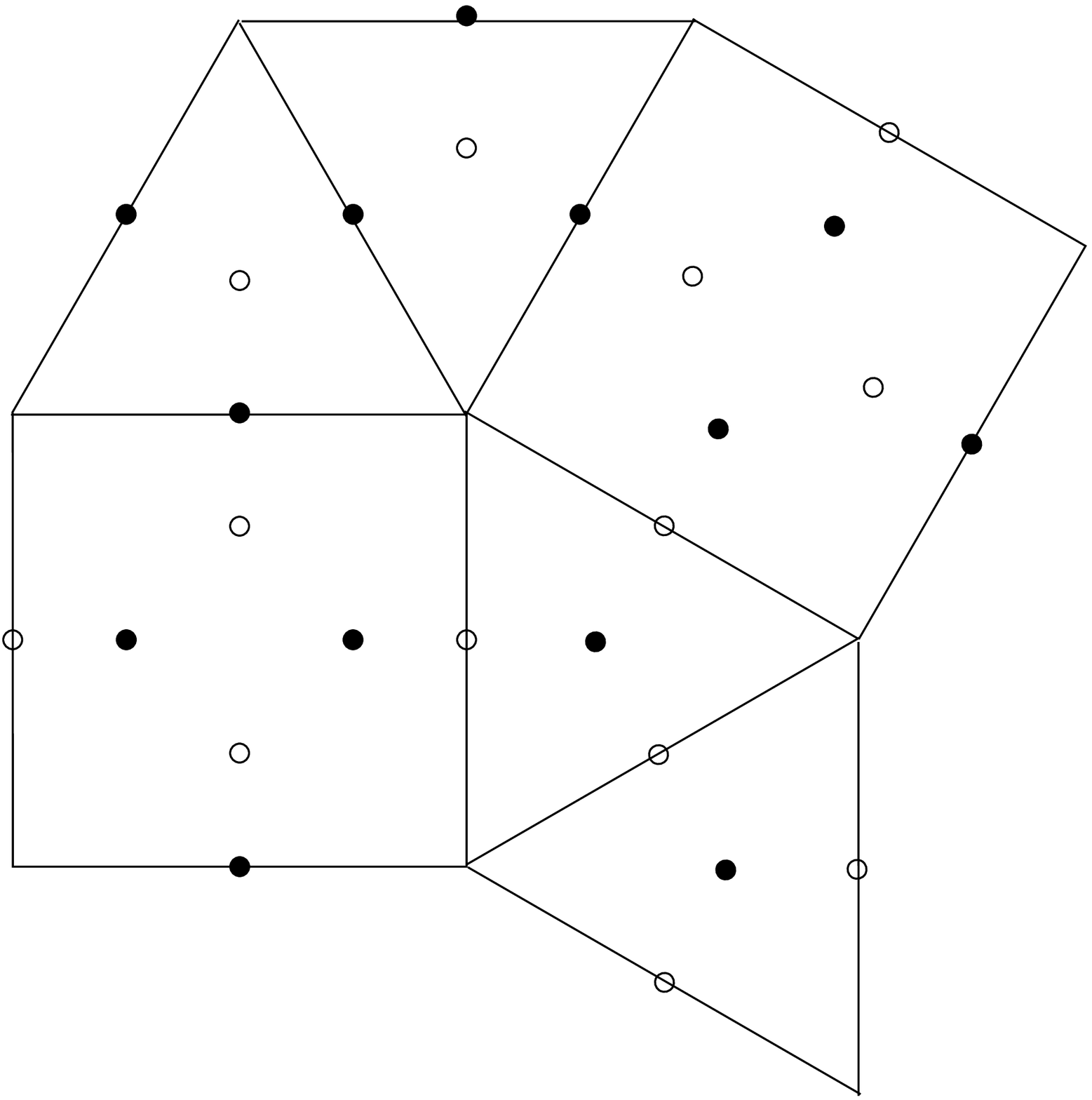,height=1.6in}\\
(c) & (d) \\
\end{tabular}
\caption{Four simple periodic square/triangle tilings of
the plane which lead to the TCP structures named
(a) Z; (b) A15; (c) $\sigma$; and (d) H.}
\label{fig:tilings1}
\end{center} \end{figure}

Sullivan~\cite{Sullivan00} has formalized a construction suggested
by Frank and Kaspar for mixing the basic TCP structures.
Start with any tiling of the plane by copies of an equilateral triangle 
  and a square, like one of the four shown in \figr{tilings1}.
Suppose the side length of the square and triangle is $4$.
Mark black and white dots on the tilings as shown.
(The dots are at edge midpoints, at triangle centers, and at distance $1$
from the sides of the sqaures.)
Then the vertices of the corresponding TCP structure are
at heights $4k-1$ above the black dots,
at heights $4k+1$ above the white dots,
and at heights $4k$ and $4k+2$ above the vertices
of the square/triangle tiling.
(Here $k$ ranges over all integers.)

Again, in each case the TCP triangulation is simply the Delaunay triangulation
of this periodic point set.  See \figr{all_cons}.
The triangulations constructed in this way are all combinations
of the A15 and Z structures.
A variant of this construction~\cite{Sullivan00}
builds combinations of the Z and C15
structures, again starting from an arbitrary square/triangle tiling.

Especially in the mixed structures like $\sigma$ and H, the particular
  geometry we have described here may differ slightly from that found
  in the actual crystals with the same combinatorics.
Presumably, these slight adjustments do not affect
  the shapes of the tetrahedra very much.
The quality figures we present below are measured using the
  exact geometry we have just described.

Sullivan's original interest in these structures was for the
mathematical study of foam geometry.  The Kelvin problem
asks for the most efficient partition of space into unit-volume
cells, that is, for the partition with least surface area.
Lord Kelvin's suggested solution was a slightly relaxed form of the 
BCC voronoi cells (truncated octahedra).  But in 1994, Weaire
and Phelan~\cite{WPh94} discovered that a relaxed form of
the voronoi cells for the TCP structure A15 is more efficient than
the Kelvin foam~\cite{KS96}.

It is perhaps not surprising that TCP structures are related to foams:
Plateau's rules for singularities in soap films minimizing their surface area
imply that a foam is combinatorially dual to some triangulation,
preferably one with nearly regular tetrahedra.
It is an interesting question whether any triangulation meeting
the combinatorial definition of TCP can be built with tetrahedra
close to regular, but certainly for the known TCP structures
this seems always to be the case.  Thus our acute triangulations
arising from this construction have high quality by almost all measures.

\subsection{Icosahedral Construction of the Z Structure}
\begin{figure}[htb] \begin{center}
\psfig{figure=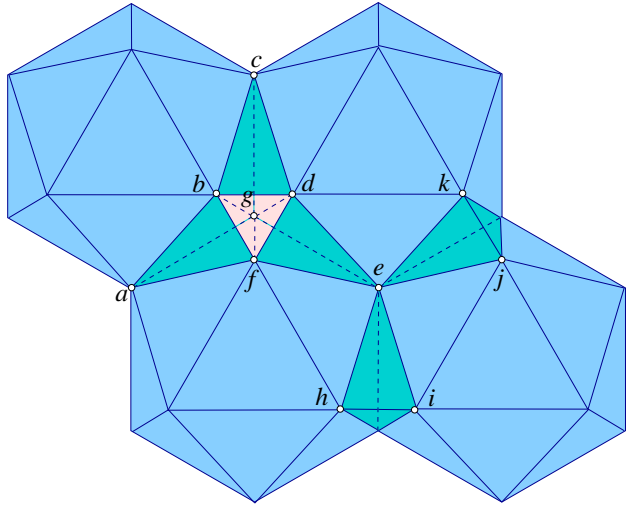,width=.5\textwidth}
\caption{Icosahedron construction.}
\label{fig:cons_ico}
\end{center} \end{figure}

An alternate construction for the TCP Z structure is
  inspired by the work of Field~\cite{Field86}.
His tilings involved right-angled tetrahedra,
  but by selectively adjusting the point set,
  we obtain a tiling with only acute tetrahedra. 
A regular icosahedron can be subdivided into 20 acute (and nearly
  regular) tetrahedra simply by coning to the center point.
We place icosahedra in a hexagonal lattice in the plane,
  each in the same orientation, touching edge to edge, as in \figr{cons_ico}.

\begin{figure}[htb] \begin{center}
\begin{tabular}{c c}
\psfig{figure=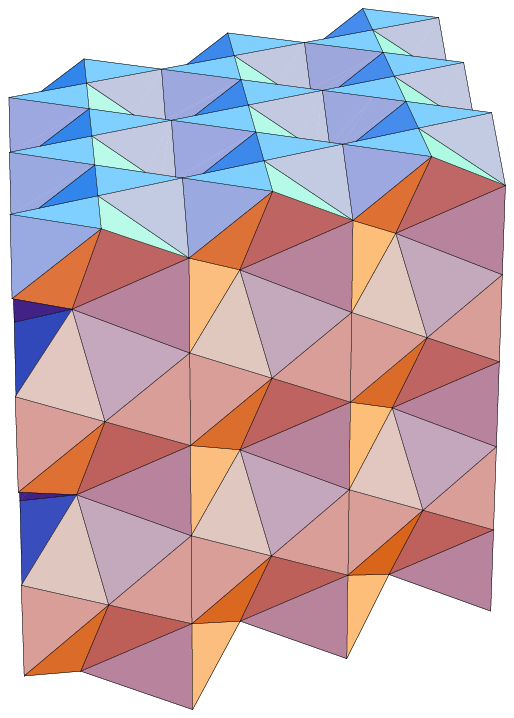,height=2in} &
\psfig{figure=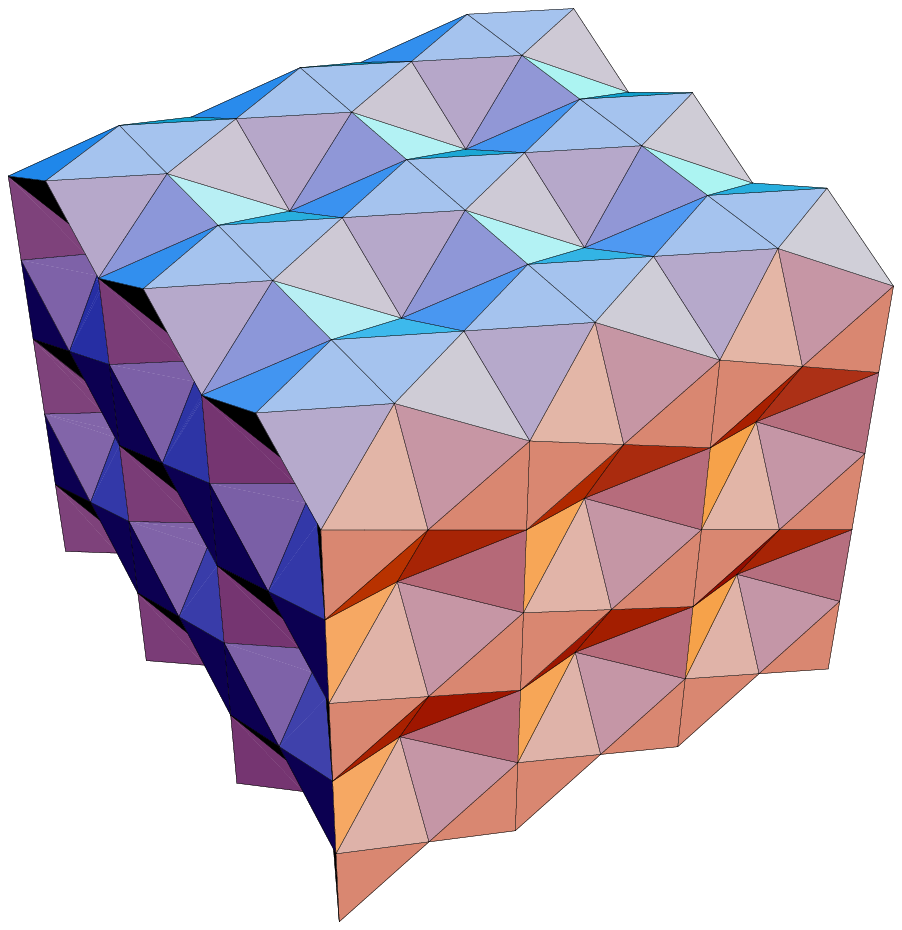,height=2in} \\
(a) & (b) \\
\psfig{figure=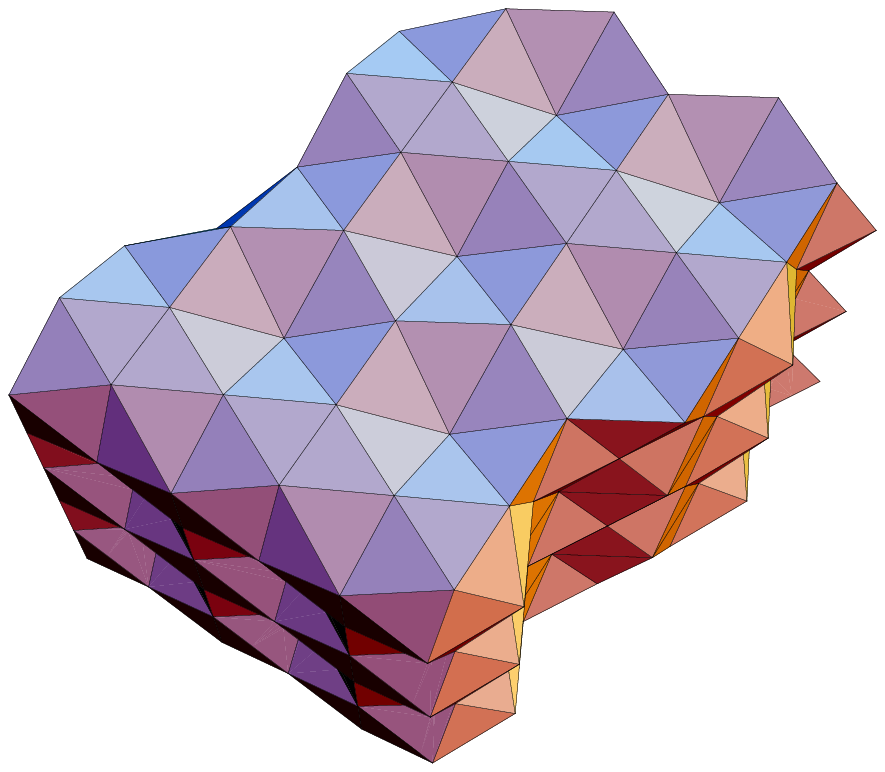,height=2.2in} &
\psfig{figure=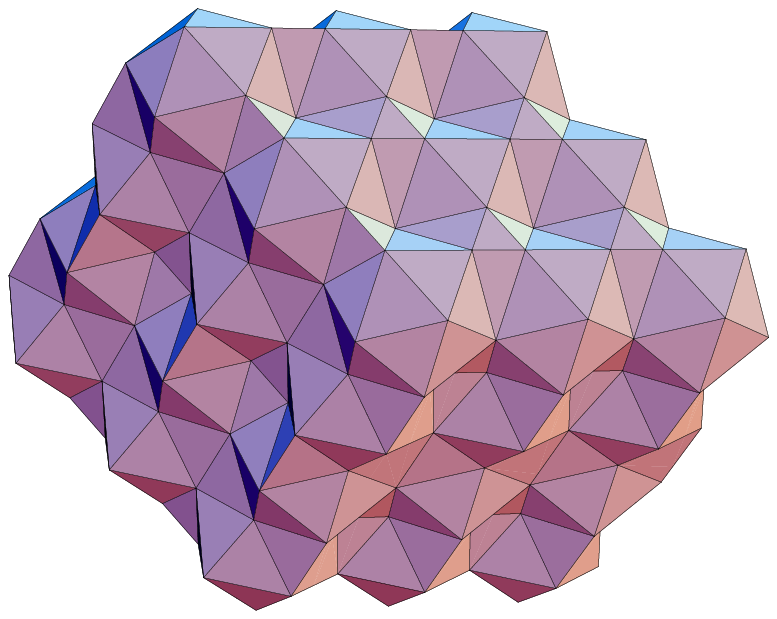,height=2.4in}\\
(c) & (d) \\
\end{tabular}
\caption{Acute triangulations filling space.
(a) The TCP structure Z (from a triangle tiling).
(b) The TCP structure A15 (from a square tiling).
(c) The TCP structure $\sigma$, a mixture of A15 and Z.
(d) Icosahedron construction of Figure \ref{fig:cons_ico}.} 
\label{fig:all_cons}
\end{center} \end{figure}

This layer then gets repeated vertically, with each icosahedron
sharing a horizontal face with the ones just above and below it.
Our point set is then the vertices and centers of all the icosahedra.
Its Delaunay triangulation, shown in \figr{all_cons}(d),
is combinatorially the TCP Z structure, but
with slightly different geometry than that constructed before.
The horizontal faces (seen head-on as equilateral
triangles in \figr{cons_ico}) are shared by two icosahedra.
Each other face separates an icosahedron from one of the
four types of Delaunay tetrahedra that fill the gaps.
There are two type of gaps. The deeper gaps are defined by 
the points $a, b, c, d, e, f, g,$ and $g'$ (opposite to $g$) and
are filled with two type of tetrahedra, e.g., $bdfg$ and $abfg$.
The shallower gaps are defined by the points 
$d,f,h,i,j,k,e,$ and $e'$ (opposite of $e$) and are filled
with two type of tetrahedra, e.g., $ee'df$ and $ee'fh$.

\subsection{An Acute Triangulation of a Single Slab}

The acute triangulations we have described so far, though periodic,
 do not have any planar boundaries within them.
Here we describe an acute triangulation of a slab (which can of
 course be repeated to fill all of space).
We view this as partial progress towards the problem of
 triangulating an arbitrary domain, although it seems much harder
 to find an acute triangulation of a cube or even an infinite square prism.
We triangulate the bottom half of the slab in the following eight steps.
Let $h$ be the height of the slab and $\gamma = h/14.2$.
\begin{figure}[tbh]
\begin{center}
\begin{tabular}{c c c}
\psfig{figure=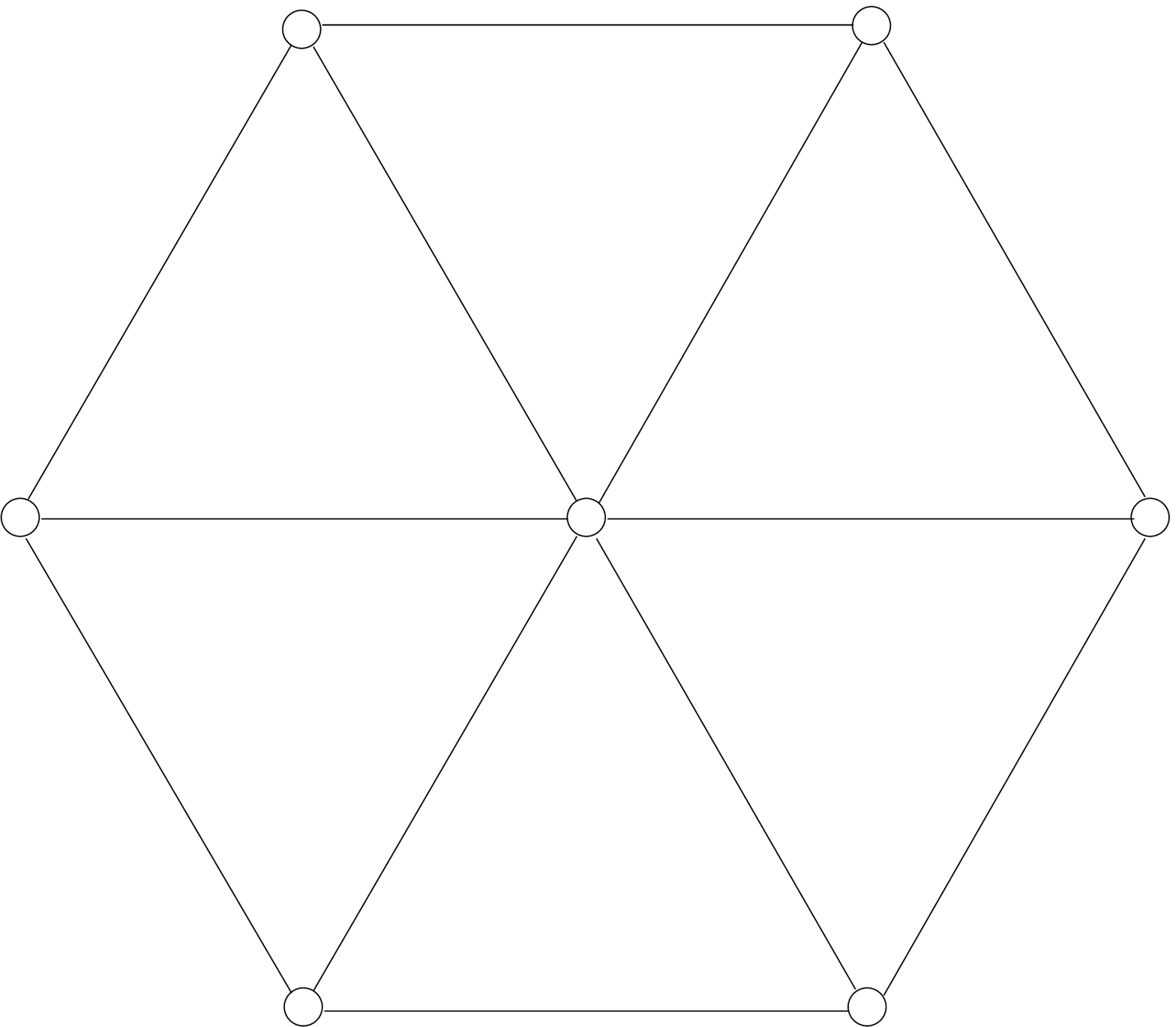,width=.26\textwidth} &
\psfig{figure=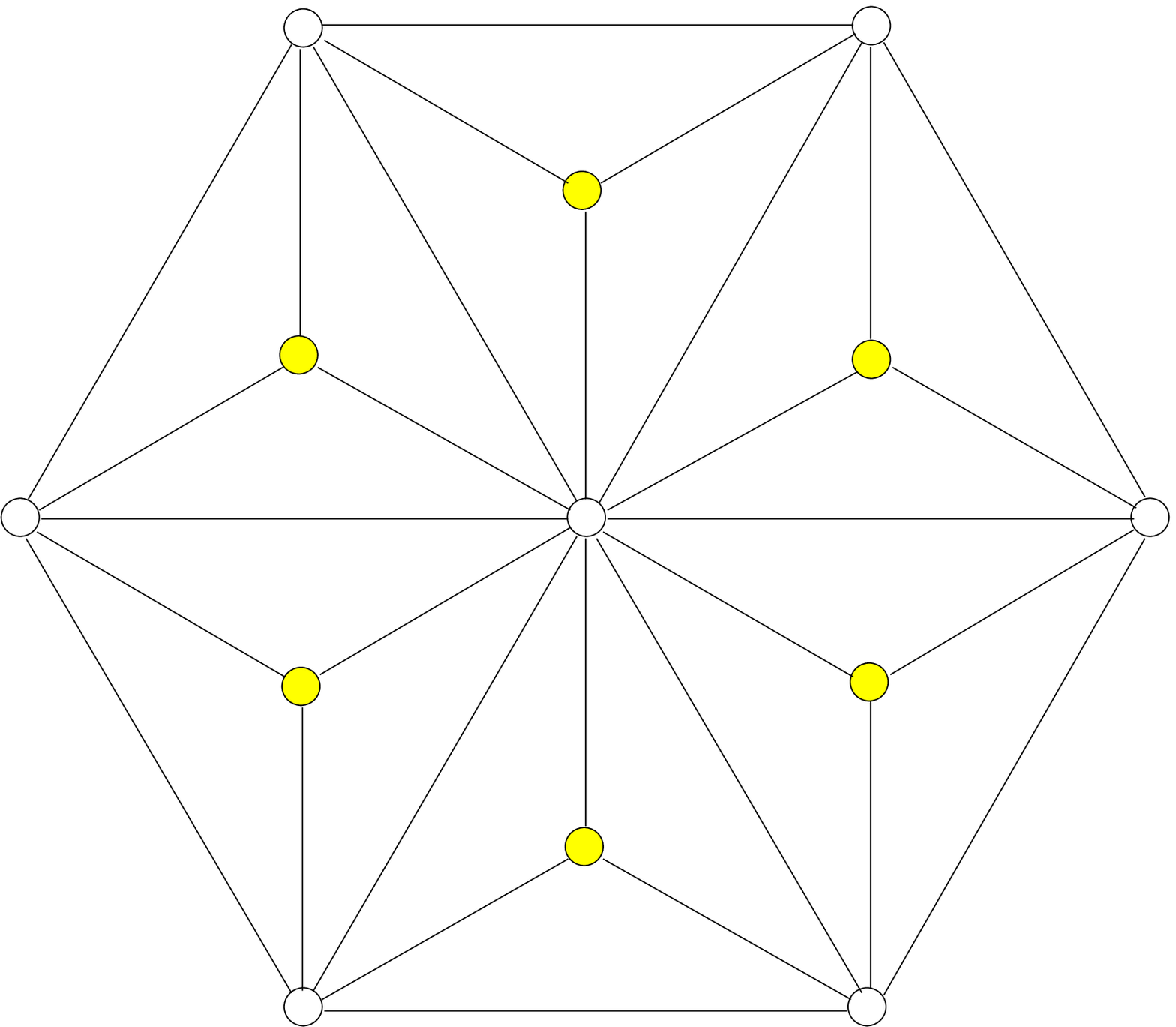,width=.26\textwidth} &
\psfig{figure=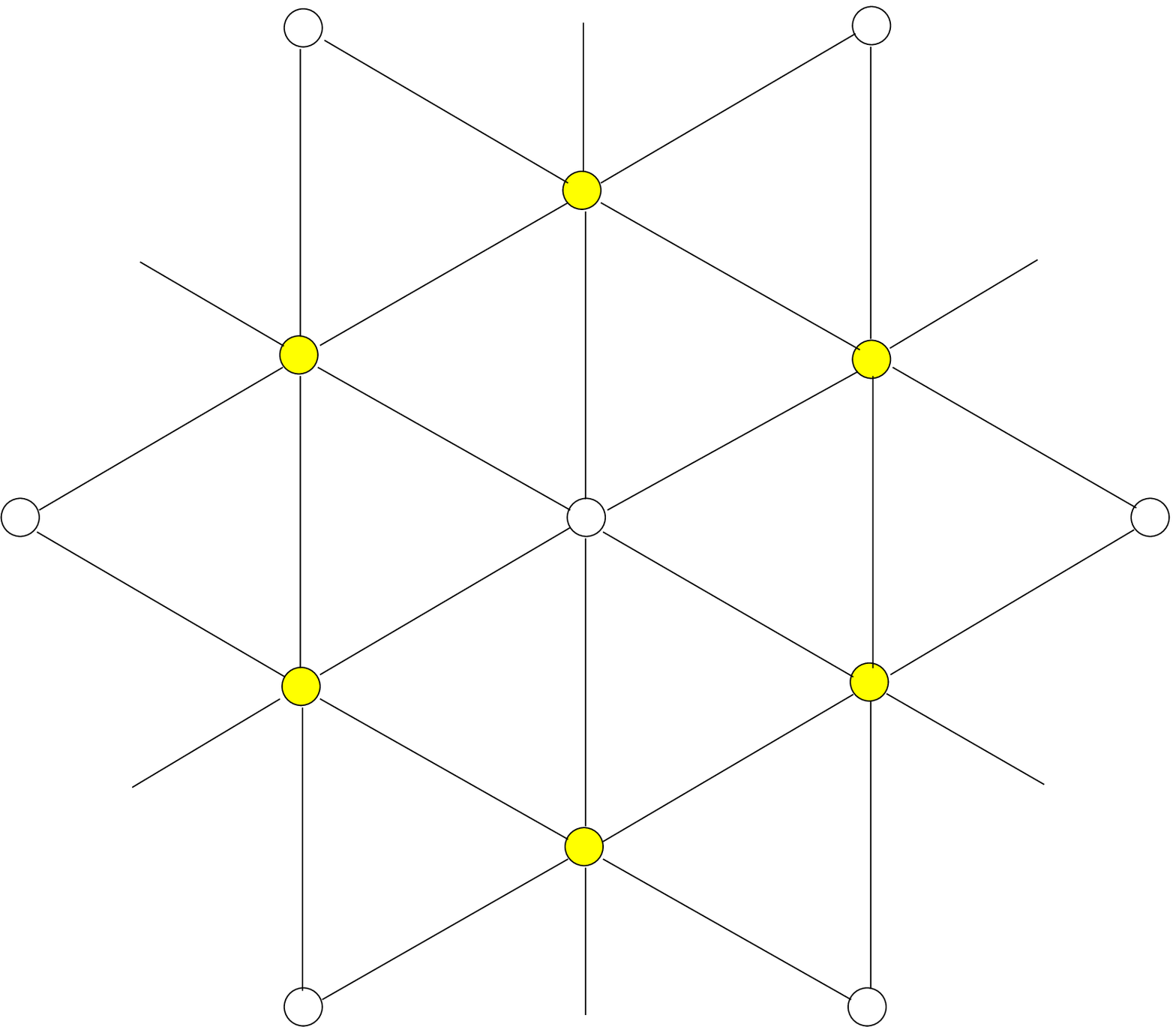,width=.26\textwidth} \\
(a) & (b) & (c) \\ \ \\
\psfig{figure=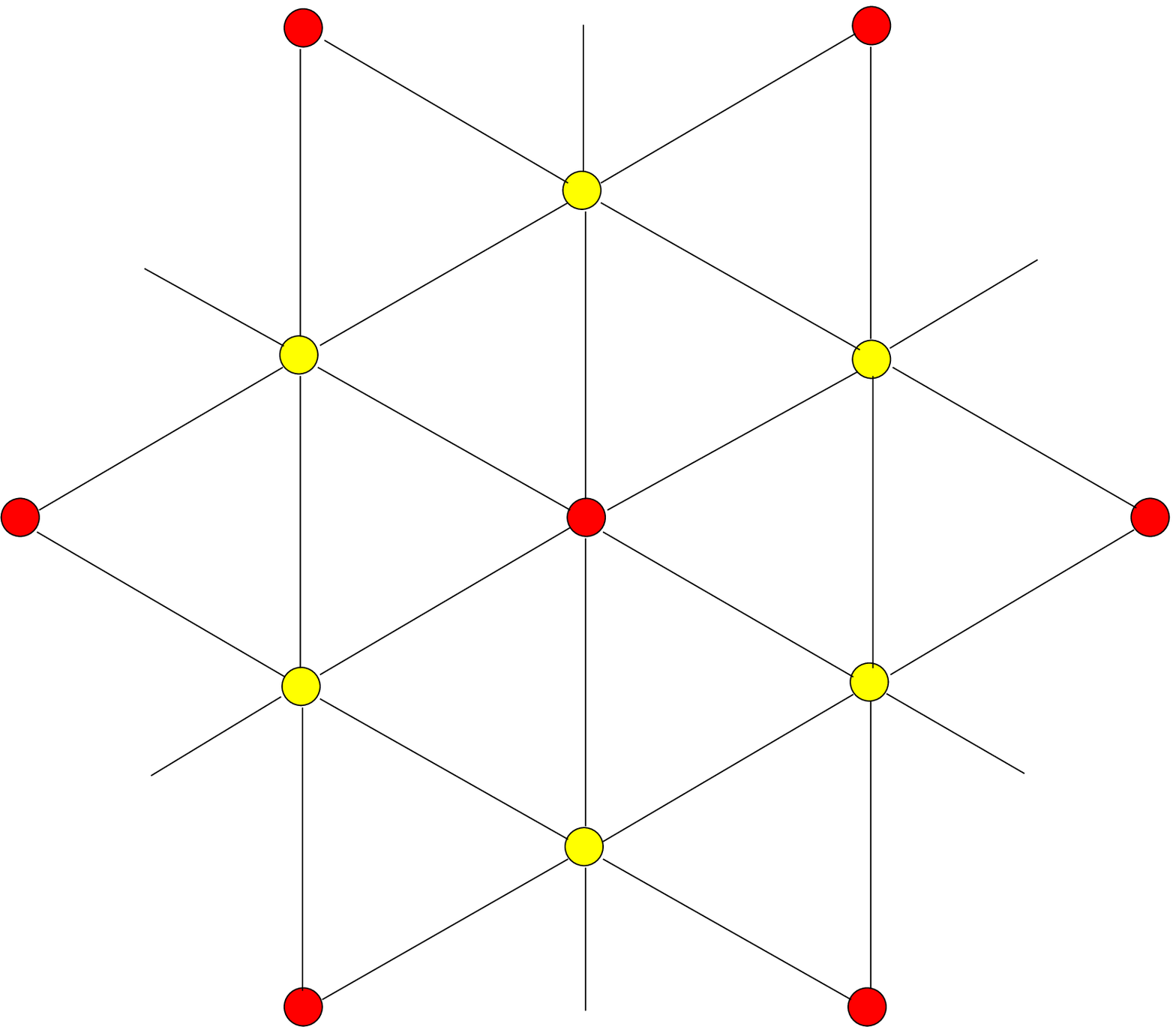,width=.26\textwidth} &
\psfig{figure=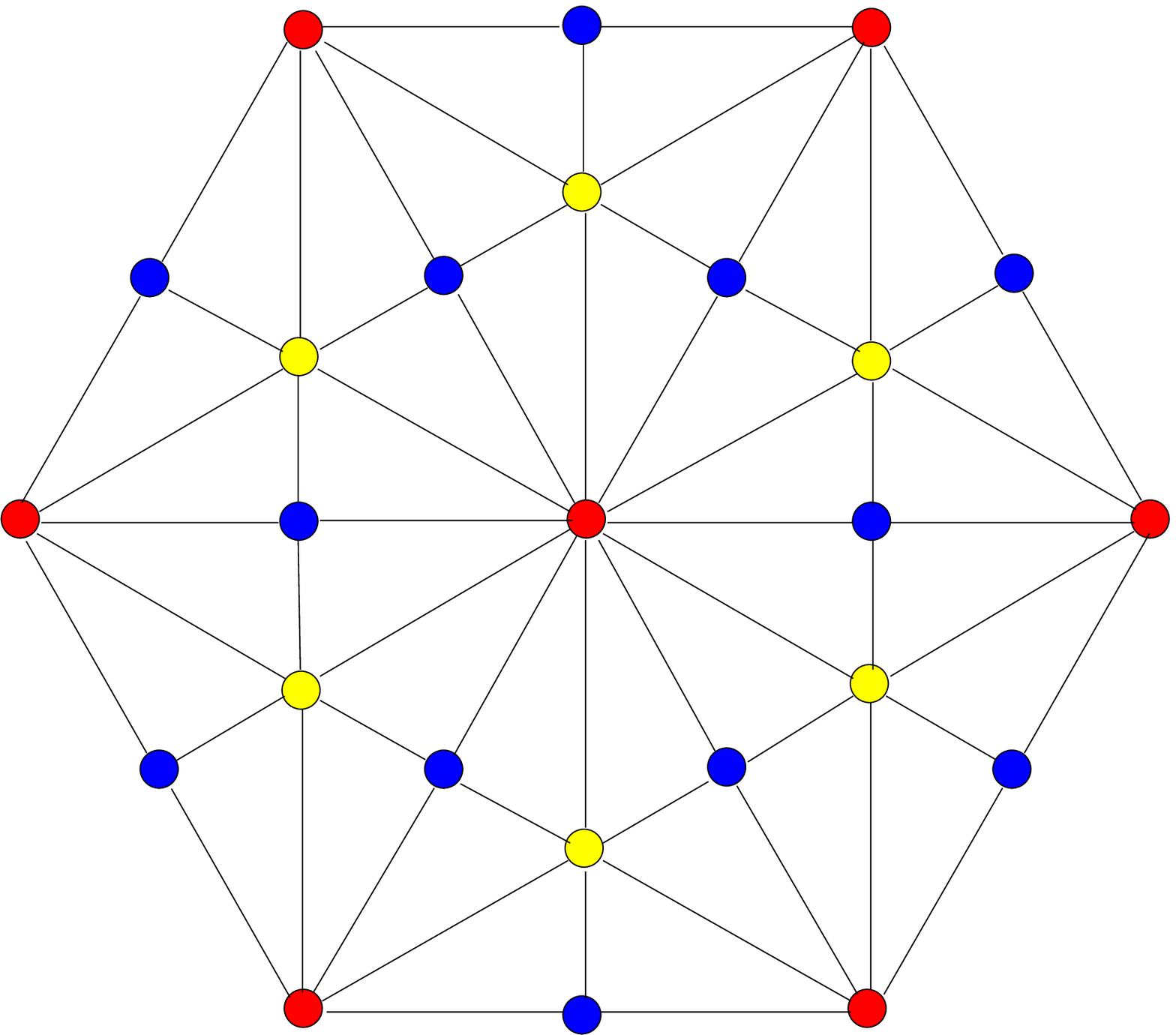,width=.26\textwidth} &
\psfig{figure=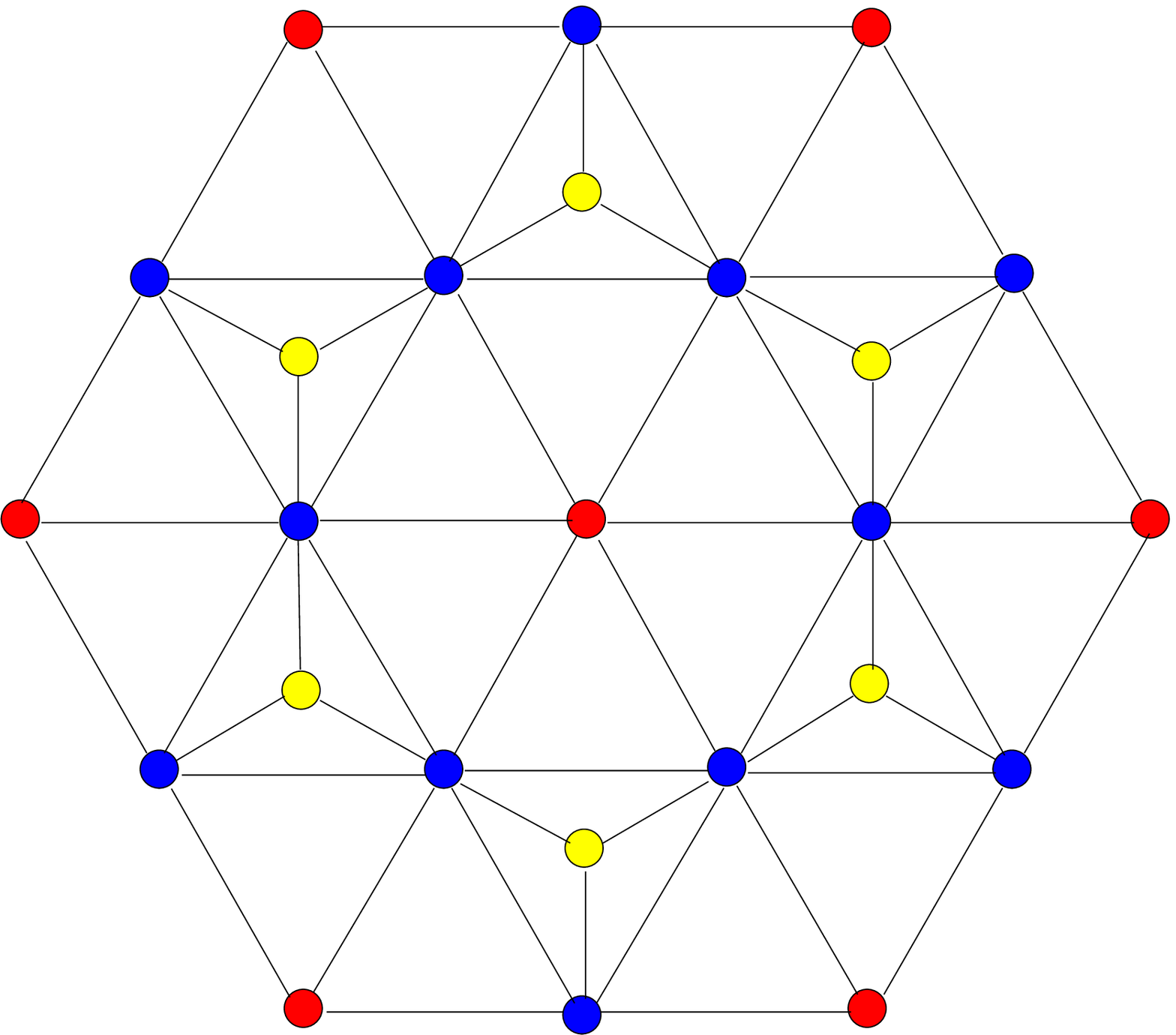,width=.26\textwidth} \\
(d) & (e) & (f) \\ \ \\
\end{tabular}
\begin{tabular}{c c}
\psfig{figure=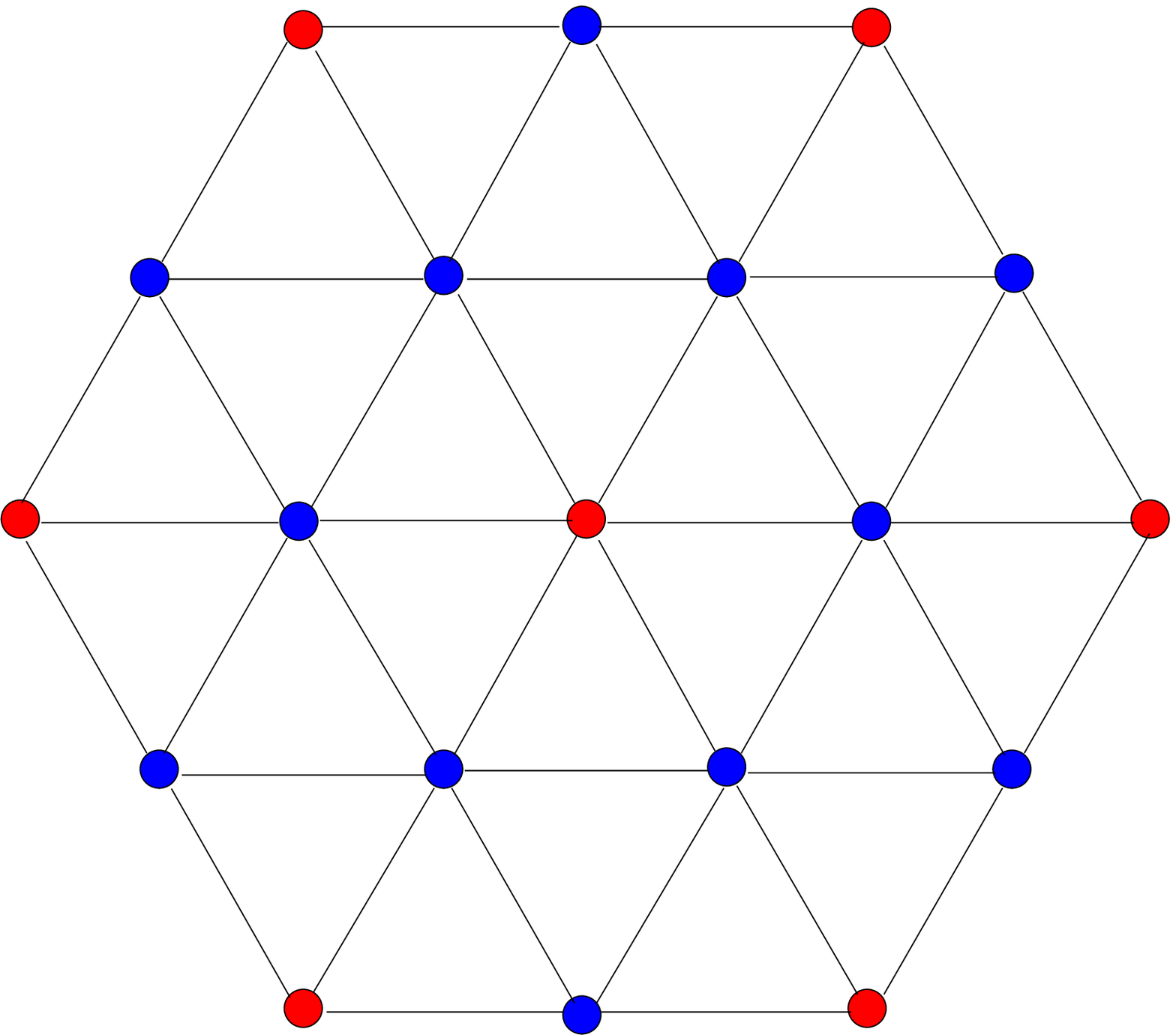,width=.26\textwidth} &
\psfig{figure=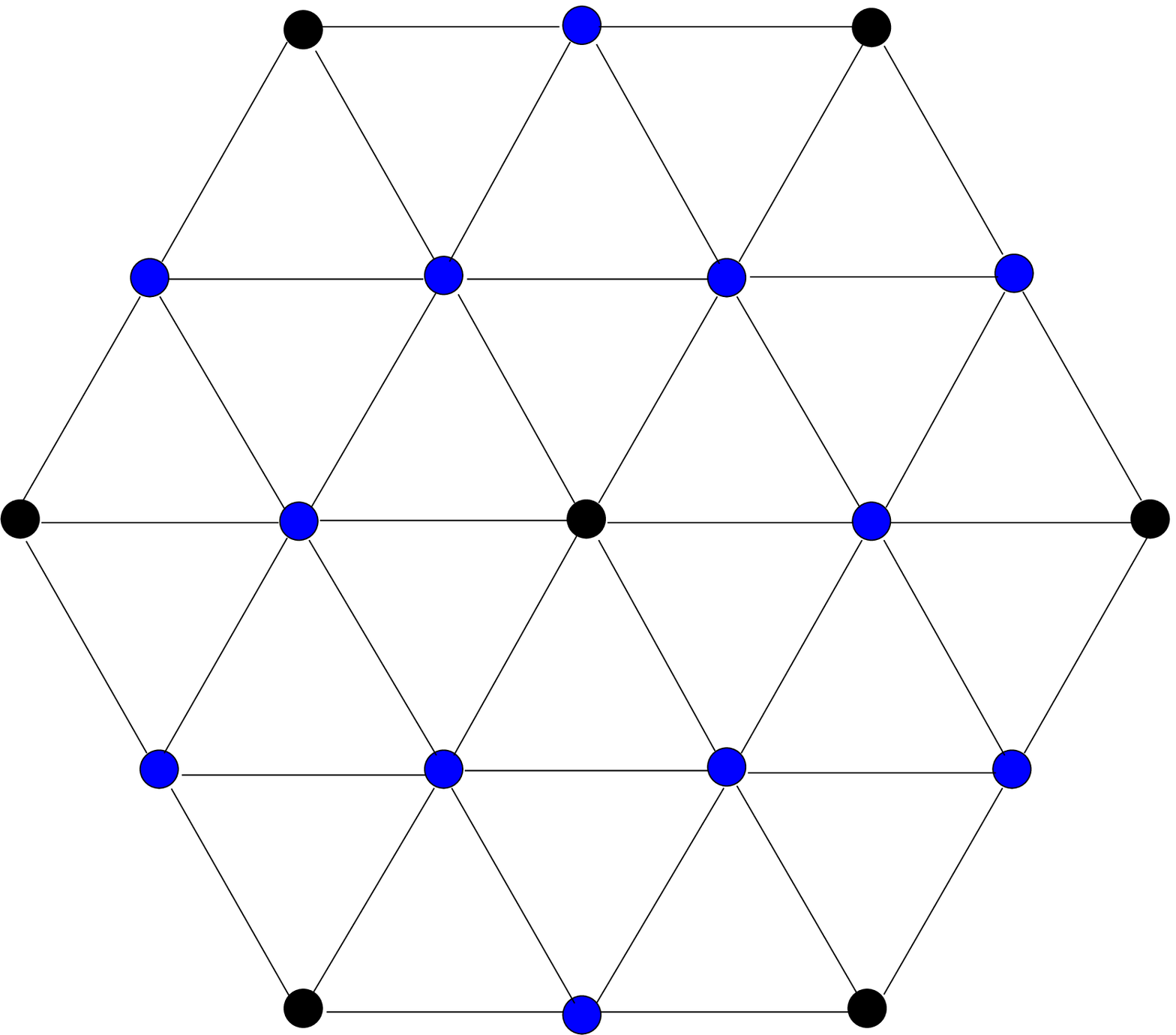,width=.26\textwidth} \\
(g) & (h)\\
\end{tabular}
\end{center}
\caption{Eight steps in filling a slab with acute tetrahedra.
The nodes in the base plane are colored white; successive layers
above that plane are then colored yellow, red, blue and black, in order.}
\label{fig:slab}
\end{figure}
\begin{enumerate}

\item Start with a grid of equilateral triangles of side length
      $6\gamma$ on the base plane, as in \figr{slab}(a). 

\item Place a near-regular tetrahedron (with height $4\gamma$) over
  each triangle, as in \figr{slab}(b). 

\item Add a tetrahedron in the gap between each pair of adjacent tetrahedra,
      as in \figr{slab}(c). The resulting surface has deep hexagonal dimples
      at the original vertices in the base plane.

\item Add six tetrahedra in each dimple, 
      each with one vertex on the starting plane, two vertices at
      height $4\gamma$, and one new vertex at height $4.6\gamma$ over the
      starting vertex, as in \figr{slab}(d). 
      Now we have a surface with shallow hexagonal bumps.

\item Place a vertex at height at $7.1\gamma$ over the midpoint of the
      edge between each pair of adjacent bumps.
      Each such vertex and edge form a vertical triangle; let this separate
      two new tetrahedra whose fourth vertices are the
      the two nearby bump vertices, as in \figr{slab}(e). 
      The surface is now covered by tall diamond-shaped bumps. 

\item Place a tetrahedron between each adjacent pair of bumps,
      as in \figr{slab}(f). 
      We now have an alternating grid of medium-depth six-sided holes
      (over each of the shallow hexagonal bumps) and deep tetrahedral holes
      (over the  points where three of the shallow hexagonal bumps meet).

\item Fill each tetrahedral hole, to form a surface alternating
      between six-sided holes and flat triangles, as in \figr{slab}(g). 

\item Place six tetrahedra into each medium-height hexagonal hole to
      turn it into a medium-height hexagonal bump, as in \figr{slab}(h). 
      In order to make the bumps equal to the holes, the height of the new
vertices is chosen as $[2 (7.1-4.6)+4.6]\gamma = 9.6 \gamma$.

\end{enumerate}

To complete the triangulation of the slab, we now repeat the first
seven steps in reverse order. 

\bigskip
Any of the constructions given in this section serves to prove
  our main result:
\begin{theorem}
It is possible to tile three-dimensional Euclidean space with acute tetrahedra.
\qed
\end{theorem}

\section{Evaluation of the constructions} \label{sec:eval}

Our constructions use tetrahedra of quite good quality
  (as summarized in Table~\ref{table:quality}) 
  and so they are quite suitable for mesh generation.
\begin{table}[bth]
\begin{center}
\begin{tabular}{|l|r|r|r|r|r|r|}
\hline 
&
\multicolumn{2}{c|}
{\begin{tabular}{c} radius-edge \\ ratio \end{tabular}} &
\multicolumn{2}{c|}
{\begin{tabular}{c} smallest \\ dihedral angle \end{tabular}} &
\multicolumn{2}{c|}
{\begin{tabular}{c} largest \\ dihedral angle \end{tabular}} \\
\hline
Construction & \hspace{.1in} min & \hspace{.1in} max & \hspace{.2in}min 
             & \hspace{.2in}max & \hspace{.2in}min & \hspace{.2in}max \\
\hline
\hline
TCP Z from triangle tiling & .651 & .737 & 53.13 & 67.37 & 73.89 & 77.07 \\
TCP A15 from square tiling & .645 & .707 & 53.13 & 67.79 & 73.39 & 78.46 \\
TCP $\sigma$ & .645 & .737 & 53.13 & 67.79 & 73.39 & 78.46 \\
TCP C15 & .612 & .711 & 60\hspace{.19in} & 70.52 & 70.52 & 74.20 \\
TCP Z from icosahedra & .629 & 1.000 & 41.81 & 69.09 & 71.99 & 83.62 \\
Slab & .636 & .938 & 46.83 & 67.88 & 74.39 & 87.70 \\
\hline
\hline
Sommerville I & 1.118 & 1.118 & 45\hspace{.19in} & 45\hspace{.19in} & 90\hspace{.19in} & 90\hspace{.19in} \\
Sommerville II & .645 & .645 & 60\hspace{.19in} & 60\hspace{.19in} & 90\hspace{.19in} & 90\hspace{.19in} \\
Sommerville III & .866 & .866 & 45\hspace{.19in} & 45\hspace{.19in} & 120\hspace{.19in} & 120\hspace{.19in} \\
Sommerville IV & 1.581 & 1.581 & 30\hspace{.19in} & 30\hspace{.19in} & 131.81 & 131.81\\
Cube V & .612 & .866 & 54.73 & 70.53 & 70.53 & 90\hspace{.19in} \\
Cube VI & .866 & .866 & 45\hspace{.19in} & 45\hspace{.19in} & 90\hspace{.19in} & 90\hspace{.19in} \\
\hline
\hline
Regular tetrahedron & .612 & .612 & 70.53 &  70.53 & 70.53 & 70.53 \\
Cube corner & .866 & .866 & 54.73 & 54.73 & 90\hspace{.19in} & 90\hspace{.19in} \\
\hline
\end{tabular}
\caption{The quality of the tetrahedra in our constructions
(and of the regular tetrahedron) can be measured in terms of the 
radius-to-shortest-edge ratio and the maximum dihedral angles.}
\end{center}
\label{table:quality}
\end{table}

There are still some challenges left to make use of these 
 tilings in real-life meshing techniques.
A strategy is required to
 fit the tilings into a planar projection of the spatial domain.
Malkevitch studies this problem in \cite{Malkevitch85}.
He describes the conditions for a polygon to be tiled by squares and
  equilateral triangles. 
Also, even though one of our constructions fits between two parallel
  planes in a slab, all of them have dimples (cavities on the surface) 
  in most directrions, making them not very suitable for 
  meshing domains with flat surfaces. 

Open problems related to this work include the following
\begin{enumerate}
\item Given a point set in 3D that has an acute triangulation, 
how can we compute it?

\item Are these constructions the best possible?  For instance,
  which tiling of space with tetrahedra minimizes the maximum dihedral angle?

\item Is it possible to tile the space with congruent copies of 
  some single acute tetrahedron?

\item Is it possible to subdivide a cube (or even an acute tetrahedron) into 
  acute tetrahedra?

\item Is it possible to extend the planar acute triangulation 
  algorithm of~\cite{BernEG94} into three dimensions,
  using constructions similar to ours?
\end{enumerate}

\subsection*{Acknowledgments}
Many thanks to Jeff Erickson for discussions and 
asking how to subdivide a cube into acute tetrahedra,
Marshall Bern, Olaf Delgado, Herbert Edelsbrunner, 
and Sariel Har-Peled for discussions, 
and Damrong Guoy for his help on visualization.

This work was partially supported by NSF ITR grant DMR-0121695.
David Eppstein was supported by NSF grant CCR-9912338.
John Sullivan was also partially supported by NSF grant DMS-0071520.  
Alper \Ungor\ was also partially supported by a UIUC Computational
  Science and Engineering Fellowship.

\bibliographystyle{newabuser}
\bibliography{acute_journal}

\end{document}